# Where is the residual entropy of a glass hiding?

P.D. Gujrati

*Department of Physics, Department of Polymer Science,*

*The University of Akron, Akron, OH 44325*

(July 24, 2009)

# Abstract

We revisit the controversy, discussed recently by Goldstein in this journal [J. Chem. Phys. **128**,154510 (2008)], whether the residual entropy is real or fictional. It is shown that the residual entropy loss conjecture (ELC) at the glass transition, which results in a discontinuous entropy, (i) violates many fundamental principles of classical thermodynamics, and also (ii) contradicts some experimental facts. In particular, we discover that according to ELC (i) there will be a violation of the second law if the glass is not in equilibrium, (ii) the glass cannot be accessed in a conventional experiment, see the text, if the glass is in equilibrium, and (iii) if the discontinuity is replaced by a smooth patch, which will presumably happen for a finite system, then there is a range of temperature over which the system looses stability. Assuming, as is common in the field, that glasses are in internal equilibrium, we show that the continuity of enthalpy and volume at the glass transition require the continuity of the Gibbs free energy and the entropy, which contradicts ELC. It is then argued that ELC is founded on an incorrect understanding of what it means for a glass to be kinetically trapped in a basin and of the concept of probability and entropy. Once this misunderstanding is corrected in our approach by the proper identification of entropy as the *ensemble* entropy in accordance with the principle of reproducibility (see Sect. II), it follows immediately that the residual entropy does not disappear in a kinetically frozen glassy state and all the violations of thermodynamics disappear. We discuss why the *temporal* definition of entropy does not make sense for glasses; in particular it is not unique for finite times. We discuss the issue of ergodicity and show that there is no loss of ergodicity if we properly identify the conditions for temporal evolution. We also discuss the role of causality and show that it is intact. We therefore suggest that our approach and our formulation of the entropy as an *average* over many *independent* samples, so that it gives the experimentally measured entropy, provides the correct interpretation of the residual entropy, clarifies ergodicity and causality and restores our faith in classical thermodynamics.

## I. INTRODUCTION

### A. Entropy Loss Conjecture

In a recent paper, Goldstein [1, see also the comment on this paper by Gupta and Mauro [2] and the work by Kozliak and Lambert [3], and Nemilov [4]], has argued that if a supercooled liquid containing $N$ particles freezes into one of exponentially large $N_b$ basins in the potential energy landscape [5, 6] at the experimental glass transition temperature $T_g$, and if this results in a loss of configurational entropy [2, and references in [1, 2]] from

$$S_R \equiv \ln N_b \tag{1}$$

(known as the *residual entropy*) to 0 (we set the Boltzmann constant equal to 1), then this drop in entropy will violate the second law [7] of thermodynamics. The existence of a non-zero residual entropy does not violate Nernst's postulate, as the latter is applicable only to equilibrium states [8, Sect. 64]. Moreover, the observation of residual entropy is very common in Nature [8, Sect. 64]. Indeed, Tolman [9, Sect. 137] devotes an entire section on this issue for crystals in his seminal work, while Sethna provides an illuminating discussion for glasses [10, Sect. 5.2.2]. In addition, the existence of the residual entropy has been demonstrated rigorously for a very general spin model by Chow and Wu [11, see references in this work for other cases where the residual entropy is shown to exist rigorously]. The numerical simulation carried out by Bowles and Speedy for glassy dimers [12] also supports the existence of a residual entropy. Thus, it appears that the support in favor of the residual entropy is quite strong. However, the situation has been challenged, more vigorously recently [13] because it is argued that the entropy cannot be really measured correctly in the glass transition region, where irreversibility comes into play. While the idea that the irreversibility raises some concern about the inferred values of the entropy is certainly justified, the main point is not whether irreversibility is present. Rather, the question should be: how much of an error does this create in the inferred values of the entropy? As we will see later in Sect. I C, various estimates support an error that does not exceed more than 5%. Thus, there is no harm in our neglecting the irreversible contribution to the entropy. It is against this background that we wish to revisit the issue of the residual entropy in glasses; see Sect. I C for additional support. For residual entropies in crystals, we ask the reader to consult [1, 3, 9]. A good historical discussion of the residual entropy for crystals can be found in [9];

for glasses, see [1, 2, (b)].

The situation with a discontinuous entropy loss is schematically shown in Fig. 1, where the blue curves show the entropies of the supercooled liquid (SCL) and the glass (GL), and the dashed green vertical jump shows the discontinuity at the glass transition at $T_\text{g}$. For a finite system, this discontinuity will be replaced by a continuous piece shown by the dashed red curve, which we will discuss later. For the moment, we are interested in the discontinuity associated with a macroscopic system. The discontinuous drop in the entropy at the transition, as the temperature is lowered, is known as the *entropy loss conjecture* (ELC). This conjecture has received a lot of attention recently [13] by its proponents and opponents, and the situation is not very clear. Some of the notable references are [1–4, 14–18].

Gupta and Mauro [2] consider the glass to freeze in one of $N_\text{MB}$ metabasins, rather than in a basin, where a metabasin by definition contains several basins. The sum of all basins in all metabasins is $N_\text{b}$. The drop $\Delta S(T_\text{g})$ in the entropy at $T_\text{g}$, which is given by

$$\Delta S(T_\text{g}) \equiv S_\text{SCL}(T_\text{g}) - S_\text{GL}(T_\text{g}) > 0,$$

represents somewhat of a smaller discontinuity than $S_\text{R}$ when the glass freezes in a metabasin. Thus, ELC should be thought of as a conjecture about the presence of a *non-zero* discontinuity $\Delta S(T_\text{g})$; its actual magnitude is not important as long as it is non-zero, but is most certainly larger than about 5% of $S_\text{R}$ for reasons to be explained in Sect. I C. We will usually take $\Delta S(T_\text{g}) = S_\text{R}$ for reasons that we explain in Sect. I B.

It is agreed by all, and this is also supported by experiments, that the enthalpy $H$ and the volume $V$ show no discontinuity at the glass transition. Consequently, the entropy discontinuity gives rise to a discontinuous jump downwards in the Gibbs free energy $G = H - TS$ [2] in the amount of $T_\text{g}\Delta S(T_\text{g})$, as shown by the blue curves and the green discontinuity in Fig. 2.

Gupta and Mauro [2] make the following observations to support ELC:

ELC1  GL is confined to one of the many metabasins at $T_\text{g}$.

ELC2  SCL and GL are two different macrostates at $T_\text{g}$.

ELC3  GL is not an equilibrium state, but SCL is (within the restricted framework in which crystallization in not allowed). GL is a broken-ergodic state [19] in which slow processes (time scale $t_{\text{s}}$) remain frozen, but fast processes (time scale $t_{\text{f}}$) have equilibrated, while SCL is ergodic in which both processes have equilibrated. Despite this, GL is considered to be under equilibrium (which we will call internal equilibrium; see below in Sect. II) within its metabasin.

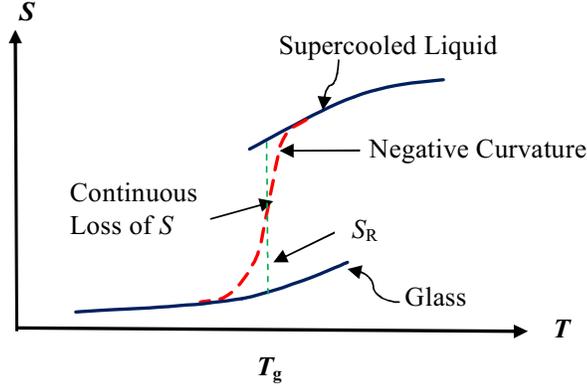

FIG. 1: Schematic form of an abrupt entropy drop (dashed green vertical line) of $S_{\text{R}}$ at the glass transition $T_{\text{g}}$, and its schematic continuous variation (dashed red curve) for a finite system that avoids the sharp discontinuity. The entropy of the supercooled liquid and of the glass is shown by blue curves. Note the curvature of the entropy is negative near the top of the continuous red patch, which causes the violation of stability, as discussed later.

ELC4 The transition to GL is a non-spontaneous process, which is not reversible because GL is a non-equilibrium state.

ELC5 The glass transition is determined by the experimental observation time $\tau$, and occurs when
$$t_{\text{s}}(T_{\text{g}}, P) \equiv \tau; \qquad (2)$$
accordingly, the following inequalities must always be satisfied
$$t_{\text{f}}(T, P) << \tau \leq t_{\text{s}}(T, P). \qquad (3)$$

We have modified the above inequalities from the one quoted by Gupta and Mauro [2, (a), Eq.(4)] to ensure that (2) can actually be satisfied.

ELC6 A non-spontaneous process can be accomplished only by performing work on the system [2, (b) Sect. 5.1].

ELC7 A less-constrained state (such as an equilibrium state) cannot return to a constrained state (such as a non-equilibrium state) without some external intervention [2, (b) Sect. 5.1].

ELC8 The entropy of GL increases as it relaxes towards the equilibrium SCL.

ELC9 The Gibbs entropy cannot be extended to describe glasses.

It is evident that the controversy centers around certain critical concepts, which include the slowly varying macrostates, entropy, irreversibility, loss of ergodicity, thermodynamic potentials and approach to equilibrium. While irreversible thermodynamics is a standard topic dealt in many textbooks, the general formalism is not very useful in making any quantitative prediction. Thus, one invariably uses the idea of *local equilibrium*, see for example [20], that allows us to apply the standard ideas of equilibrium thermodynamics locally. As long as the system is under ordinary conditions, such as the absence of turbulence, the absence of chemical explosion, etc. this approach has proved quite useful and its predictions are quite reliable. All we have to do is to note its usefulness in providing a reliable description of hydrodynamics under ordinary conditions, chemical reactions, etc. Thus, our approach will be based on using the idea of local equilibrium, which we will extend to slowly-varying (in time) metastable states such as supercooled liquids and glasses in the form of internal equilibrium; see ELC3.

In a nutshell, the most forceful argument in support of ELC is the following regarding the concept of entropy. We quote Gupta and Mauro [2, (a)]:

> "We use Boltzmann's definition, which is determined by the microstates sampled or accessed by the system during the time of observation."

It follows then that once the system has been kinetically frozen in a basin for a long period of time, much longer than the period of observation, then it effectively means that there is only one "microstate," so that the Boltzmann entropy becomes zero. We again quote Gupta and Mauro [2, (a)]:

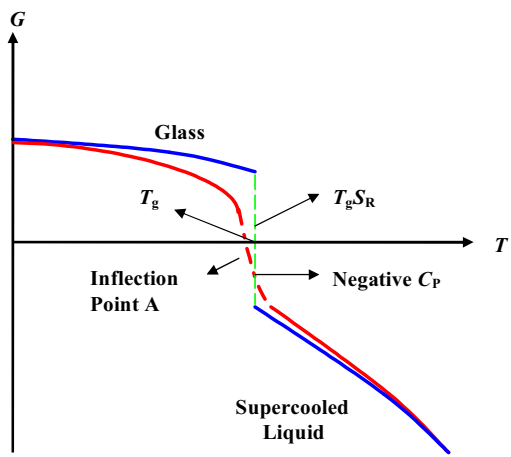

FIG. 2: Schematic form of the concave free energy $G$ as a function of the temperature $T$. The blue curve represents the Gibbs free energy for a macroscopically large system; the green discontinuity $T_g S_R$ at $T_g$ shown by the broken green portion corresponds to the residual entropy loss $S_R$. (This discontinuity will be absent if there were no entropy loss.) For finite systems, this dicontinuity will be replaced by a continuous curve shown by red, which can be conveniently divided into three different pieces: the two solid red pieces connected by a dot-dashed piece with an inflection point A. This continuous curve will eventually evolve and give rise to the green discontinuity as the size of the system becomes macroscopic ($N \to \infty$).

> "$\cdots$ at 0 K$\cdots$ the system is kinetically trapped in a single microstate. In other words, each metabasin at 0 K consists of just a single microstate. The entropy of each metabasin is zero, so the overall entropy of the system is zero. Our argument has nothing to do with the third law and is based on the fact that any transitions$\cdots$ must freeze out at low temperatures before reaching 0 K."

This seems to be at the heart of the controversy, and it is this concept of entropy and the role of kinetics that we must examine in this work. Specifically, we pose two different but relevant questions:

Q1 Is the entropy defined for a single system or is it an average property of many systems?

The answer to this question will allow us to decide if Boltzmann's entropy should be applied to a microstate or to a macrostate. We will see later that at a given instance a system will be in one microstate, while a macrostate will require considering many systems at the same instance. This explains the above wordings of the question Q1.

Q2 What role does a dynamics play in determining the value of the entropy?

For example, we wish to ask: how should we identify the entropy of $N$ non-interacting Ising spins? These spins, see Sect. IV B, have *no dynamics* so they are similar to the system confined in one of the basins. Such a situation normally occurs at absolute zero, where no classical dynamics is possible; quantum dynamics though is possible. Treating the confinement in a basin as represnting a "microstate," and applying the Boltzamnn entropy to this basin will result in a zero configurational entropy. This is consistent with the above quote. Now, different systems will be confined to different basins. If the entropy is determined by the number of these basins, then it will not be zero, even if each system remains frozen in its basin (no dynamics). Which is the correct answer?

Hopefully, by the end, we will be able to convince the reader of the correct answers to these important questions so that we will be able to provide a resolution of the controversy.

### B. Basins versus Metabasin

According to ELC1, GL becomes confined to one of exponentially large $N_{\text{MB}}$ metabasin at $T_{\text{g}}$, while SCL enjoys the exploration of the entire landscape (not containing the part describing the crystal [6], which will always be implied in this work). Thus, the glass transition at $T_{\text{g}}$ is a transition in which ergodicity is lost; see ELC3. However, it should be emphasized that the loss of ergodicity that is referred to here is limited to a *finite* period of time, while the conventional idea of ergodicity requires *infinite* time. Within the metabasin, GL is in (internal) equilibrium; see ELC3. Now, a metabasin is just like a miniature enthalpy landscape [6] as long as GL remains confined in the metabasin. The metabasin contains many basins, all different in their shapes and depths so that their enthalpy minima give rise to a broad distribution just like in the original landscape [6]. The GL in the metabasin is similar to SCL in the original landscape in that both are in equilibrium (within their appropriate landscape). If we now follow the evolution of GL in the metabasin, we expect the existence of a lower temperature at which there should occur *another loss of ergodicity*, probably due to the kinetic freezing of the secondary relaxation, the Johari-Goldstein relaxation [21], so that GL will no longer be able to explore the entire metabasin (miniature landscape), but will be confined to a subset of basins in the metabasin; this subset plays the same role as the metabasin in the context of the original basin. Such a situation must arise since within this miniature landscape, there must appear a slow dynamics, slower than $\tau$, at some $T < T_\text{g}$ that restricts GL from exploring all the basins within the metabasin; it can still probe the basins within the above-mentioned subset. This is what is expected of the time-scale of the Johari-Goldstein relaxation, which continues to grow as inverse temperature so that it will eventually exceed $\tau$ at some lower temperature. This will give rise to another loss of ergodicity with another entropy loss discontinuity below $T_\text{g}$. Indeed, there is no reason that there could not be a cascade of ergodicity losses, each with its own discontinuous entropy loss, until finally GL is trapped in a single basin. It does not appear that such a cascade is seen experimentally.

As the magnitude of the entropy discontinuity is not relevant, whether GL is confined to a metabasin or a basin is not central for ELC. Accordingly, we will assume in this work that the glass is frozen in a basin, rather than a metabasin. However, the arguments we present in this work are equally valid for the latter case; all we need to do is to replace $S_\text{R}$ by $\Delta S(T_\text{g})$. No other change is needed. Our arguments relate to the thermodynamic and experimental violations due to the presence of a discontinuous drop in the entropy, or if the drop is continuous, as may probably happen for a finite system, on the violation of stability of the system. Thus, the magnitude of the discontinuity is not relevant, provided it is larger than 5%, as discussed earlier; see also Sect. I C.

### C. Our Goal

We have already posed the two most important questions Q1 and Q2 in Sect. I A whose answers will help us with our goal of resolving the controversy. Before we get to this goal, we will carefully check the validity and mutual consistency of the above ELC claims, which have been used [2] to justify ELC; see also Nemilov [4], Gutzow and Schmelzer [16], and Odagaki and Yoshimori [22] for complementary analyses. We follow Gupta and Mauro [2] in accepting their assumption of the internal equilibrium, see ELC3 and Sect. II. We do not visit Goldstein's arguments here (for another simple argument for the violation of the second law under entropy loss, see [23, pp. 22-23 and Fig. 14a]) as they are not central to the current discussion. We follow a different line of attack. However, we also come to the same conclusion in a more direct way, as discussed later in Sect. III. Indeed, our proof is based on a direct application of ELC7. Gutzow and Schmelzer [16] have already shown by citing some earlier analysis that the irreversible contribution to the entropy in the vicinity of the glass transition is small enough to be practically neglected. Nemilov [4] comes to a similar conclusion by suggesting that the error in $S_{\text{R}}$ is no more than 5%. Jäckle [17, (a)] has also calculated the amount of uncertainty in the entropy for glycerol due to irreversibility near the glass transition, and found it to be less than 5%. Thus, there is no harm in neglecting this irreversible contribution. This suggests that the concept of internal equilibrium, see ELC3, is generally valid for typical glass forming systems. It is not surprising, therefore, that the use of internal equilibrium seems to be a standard practice at present [1, 2, 4, 14, 16, 22, and references therein]. Just to be complete, we also analyze the situation in which the abrupt discontinuity is replaced by a continuous patch as shown in Figs. (1) and (2). Within the framework of internal equilibrium, we find that the loss of entropy (abrupt or continuous) gives rise to not only various additional violations of thermodynamics but also contradicts some important experimental facts. Indeed, the standard thermodynamics must be altered due to these violations. This is a dramatic shift in the standard paradigm of using thermodynamic concepts to describe slowly-varying (in time) states of matter. Thus, *a resolution in favor of no entropy loss has to be found*, specially because the concept of using Boltzmann's entropy to a microstate or a single system [2] appears quite appealing and seductive for a glass. This has to be overcome and is also our main goal; see Q1 and Q2. A careful analysis presented here shows that there is really no violation of thermodynamics. We argue that the controversy results from an incorrect application of the concept of probability and, hence, entropy. The latter is determined not by a microstate, but by their probabilities in a macrostate, as we will elaborate later. We find that the ensemble entropy is more fundamental a quantity than the temporal entropy, not only for glasses but in general. This discussion brings in the concept of causality, see Sect. IV E, which has also been invoked [15] in support of ELC. In the process, we provide another but direct proof that the residual entropy in real glasses can in general be non-zero, as Goldstein suggests. This is also the conclusion arrived at by Nemilov [4], and Gutzow and Schmelzer [16].

### D. Outline

In the next section, we recapitulate and extend some salient principles of equilibrium thermodynamics, which we expect to be valid for slowly evolving metastable states. We provide some general arguments for their applicability, and prove one of them in Sect. III by exploiting certain experimental facts. Sect. III is an important part of the work, where we critically analyze the paradoxical consequences of ELC, which contradict not only the above principles but also experimental facts. We then use the experimentally observed *continuity* of the enthalpy and volume to show that the Gibbs free energy and entropy must also be continuous, which contradicts the basic premise of ELC. The continuous nature of entropy and free energy for a macroscopic systems that we claim should not be confused with the continuous variation due to finite size of the system; the latter will not only violate the condition of stability, but will also result in give discontinuities in the limit of large systems. Thus, we conclude that ELC is untenable. To understand the root of the controversy and to seek a resolution, we must clearly understand the concept of entropy, a probabilistic quantity, the general formalism of which is given in Sect. IV. This section covers the general formalism to discuss glasses and is of central importance. Some of our arguments are not new [8, 9, 24–29] but seem to have been either forgotten or have been overlooked by recent workers, who put an unnecessary emphasis on ergodicity and Hamiltonian mechanics or Liouville's theorem [28]. We discuss that while the concept of ergodicity may be appealing for its historical aspect, it is most certainly not very useful to study glasses whose properties are controlled by the duration of time over which they are observed. Nevertheless, we discuss ensemble and temporal averages and their applicability to slowly evolving metastable states, and the concept of ergodicity within the context of the glass transition. This discussion offers an answer for our second question Q2. We also argue in Sect. IV E that what Reiss [15, (a)] calls causality is not really violated despite his claim. Reiss's causality is nothing but the well-accepted concept of independent evolution of systems, each kinetically frozen in its own basin, in an ensemble. He wonders: how can independent systems affect each other? However, this trivial independence does not make their probabilities, which appear in the entropy, independent. Thus, there is no violation of his causality. The causality introduced by Penrose [29], however, is in our opinion the proper causality for a thermodynamic system and which remains valid in our approach. In Sect. V, we tackle the issue of the glass transition and the issue of continuity of the entropy by carefully analyzing how the residual entropy should be properly defined, ensuring its conformity with the principles laid down in Sect. V. We show that ELC results from an incorrect understanding of the concept of entropy and probability. The entropy is defined not by a single microstate, but for a macrostate. This thus answers our first question Q1. We discuss the issue of relaxation in glasses Sect. VI, and we conclude that ELC8 cannot be correct. Nernst's postulate and ergodicity are discussed in the following section, which complements the analysis of Gutzow and Schmelzer [16]. We briefly summarize our conclusions in the last section.

## II. RELEVANT THERMODYNAMIC PRINCIPLES

We will often resort to the potential energy (to be precise, enthalpy) landscape picture to facilitate our discussion as it has become a very common mode of description, but the arguments we present are general and do not depend on its validity in any way. They are based on fundamental principles of thermodynamics, which we list here not only to set the stage but also for continuity. These principles are well known in equilibrium thermodynamics. We now discuss why they should also be applicable to *slowly* evolving metastable states such as glasses in which we are interested in this work.

1. The first one is the *principle of additivity*, according to which the total entropy or other extensive quantities can be obtained by a sum over different macroscopic parts of a system. Each part must be large enough so that the usual argument that their surface effects can be neglected as thermodynamically unimportant is valid so that each part becomes *almost independent* [8].

This principle is consistent with what one must do in reality. As thermodynamics is an experimental science, it requires verification by performing the experiment many times. In other words, different parts are nothing but the preparation of the system many times under identical conditions specified by the set of (macroscopic) variables $\xi$ (which may also include its entire history to be symbolized by $\mathbf{t}$; we will assume that the observation time $\tau$ is also part of $\mathbf{t}$). Each macroscopic state (macrostate) of the system is specified by the set $\xi$ that are either fixed for the system (such as the number of particles $N$, its volume $V$, etc.) or are fixed from the surrounding medium (such as $T$, $P$, etc.). We will consider $N, T, P$ and $\mathbf{t}$ in this work as the variables to characterize the macrostate of the system. Thus, we consider a single component system, which in no way restricts the results obtained here; they can be easily generalized to more complex systems. In equilibrium, the history becomes irrelevant. However, for metastable states such as glasses, it becomes extremely important: different histories will give different macrostates. We will refer to different preparations of the system under identical macroscopic conditions (which includes **t** also) as *replicas* for brevity, and the set of all replicas as an *ensemble*. Every thermodynamic quantity must be obtained as an *average* over these replicas or macroscopic parts; see (14). At a given instant in the history, each replica will represent a particular microstate $j$ of the system, and their collection with appropriate probabilities $p_j$ represent a macrostate of the system. In equilibrium, these probabilities are history-independent. However, for metastable states, $p_j(\mathbf{t})$ are going to depend on the history **t**.

2. The second principle is the *principle of reproducibility*, according to which the ensemble average is equal to the average of the experimental values, also called the thermodynamic average.

This principle follows from the above realization that thermodynamics, whether equilibrium or time-dependent, is an experimental science and requires *several measurements* on the system to obtain reliable results. To avoid any influence of the possible changes in the system brought about by measurements, we can instead prepare a large number $N_{\mathrm{S}}$ of samples or replicas under *identical macroscopic conditions*. The replicas are otherwise *independent* of each other in that they evolve independently in time. This is consistent with the requirement that different measurements should not influence each other. As we will see in Sect. IV E, this fact has been mistakenly used to argue for the violation of causality [15]. The average over these samples of some thermodynamic quantity then determines the thermodynamic property of the system. As this replica approach will play a central role in our formalism, we state it as a fundamental axiom:

FUNDAMENTAL AXIOM *The thermodynamic behavior of a system is not the behavior of a single system, but the average behavior of a large number of independent systems, prepared identically under the same macroscopic conditions.*

Such an approach is standard in equilibrium statistical mechanics [8, 9, 23, 24], but it must also apply to systems not in equilibrium. For the latter, this averaging must be carried out by ensuring that all systems have identical history **t**. This is obviously not an issue for systems in equilibrium. We refer the reader to a great discussion about the status of statistical mechanics and its statistical nature by Tolman [9, Sect. 25]. There, Tolman clearly puts down the viewpoint of statistical mechanics as follows. We quote [9, p. 65 ]

> "The methods are essentially statistical in character and only purport to give results that may be expected on the average rather than precisely expected for any particular system.....The methods being statistical in character have to be based on some hypothesis as to *a priori* probabilities, and the hypothesis chosen is the only postulate that can be introduced without proceeding in an arbitrary manner...."

Tolman then goes on to argue that what statistical mechanics should strive for is to ensure [9, see p. 67, Sect. 25], see also Jaynes [25, last paragraph, p. 106],

> "...that the averages obtained on successive trials of the same experiment will agree with the ensemble average, thus permitting any particular individual system to exhibit a behavior in time very different from the average."

3. The third principle is the *principle of uniqueness*, which states that these thermodynamic potential relevant for a given $\xi$ is a unique (single-valued) function of the set $\xi$. In this work, the Gibbs free energy $G$ will be the relevant thermodynamic potential for $N, T, P$ and **t**.

For equilibrium states, this principle is certainly valid. We extend it now to metastable states for which different histories will give different values of the thermodynamic potential. However, we also demand that there must be a unique thermodynamic potential for a given history, otherwise different experimentalists will have no way to effectively communicate their results and no scientific investigation can be carried out. In order to define thermodynamic potentials, we require that the system be at least in internal equilibrium, called partial equilibrium by Landau and Lifshitz [8], so that we can introduce temperature, pressure, etc. for its various parts. This is similar to having local equilibrium which requires local identification of temperature, pressure, etc. so that local qualities like the entropy, energy, etc. satisfy all rules of equilibrium thermodynamics [20]. These quantities will vary with time until equilibrium with the surroundings is reached, in which case they become equal to the temperature, pressure, etc. of the surroundings. We will say that the system or each part is in *internal equilibrium*, but (complete) equilibrium (with the surroundings) has not been reached. Thus, the system will relax in time ($t > t_\text{s}$) so that its thermodynamic potential will continuously decrease until equilibrium is reached. This will be the situation with GL, which is not in equilibrium as SCL is. Here, we are talking about the equilibrium with the surroundings. During any point in its relaxation, GL has well-defined temperature, pressure, etc., at that point, though different from the surroundings. The exploration of the full phase space is not possible until $t \geq t_\text{s}$.

There will, however, be *no* relaxation for $t_\text{f}(T,P) < t < \tau$, as the system is in internal equilibrium during this time. Thus, the entropy of the system has achieved its maximum possible value *consistent* with the internal equilibrium that has occurred during this period so that the entropy can be treated as an "equilibrium" entropy associated with the internal equilibrium. (This "equilibrium" entropy will increase further only during system's relaxation for $t \geq t_\text{s}$.) Any heat exchange with the surroundings during $t_\text{f}(T,P) < t < \tau$ is related to the change in the entropy and pressure as if we are dealing with a *reversible process*

$$\Delta H = T\Delta S + V\Delta P, \tag{4}$$

where $T$ and $P$ are the temperature and pressure, respectively, of the surroundings (and not of the system or its parts); there is *no* irreversible entropic change. [See Sect. VI for a justification of (4).]

We provide a general proof of this principle for metastable supercooled liquids in points (G) and (H), Sect. III. The proof is based on continuous and single-valuedness of enthalpy and volume that has been tested in and verified by all experiments we are familiar with.

In this work, we are interested in a system that is cooled at a fixed rate $r$ from some predetermined equilibrium state at time $t = 0$. In particular, we will consider reducing the temperature by some predetermined value of the temperature interval $\Delta T > 0$ from its current value by bringing the temperature of the surroundings to $T - \Delta T$ over a period $\delta \equiv \Delta T/r$, and wait for a predetermined duration $\tau$ before reducing the temperature again. The pressure of the surroundings will be kept at some fixed $P$ during cooling. Then, the history is unique and one does not have to explicitly show the dependence on **t**, unless one wishes to follow relaxation of the system. In that case, one can simply use the time $t$ from the moment the temperature was reduced to the current value.

4. The last principle is the *principle of stability*, according to which the heat capacity, compressibility, etc. must remain non-negative for the system to remain stable.

To see this, we note that because of the absence of any irreversibility during $t_f(T,P) < t < \tau$ when the system is in internal equilibrium,

$$S = -(\partial G/\partial T)_P, \qquad (5)$$

so that

$$(\partial^2 G/\partial T^2)_P = -T(\partial S/\partial T)_P = -C_P \leq 0, \qquad (6)$$

where we have used (4). A similar condition also holds for the curvature of $G$ with respect to the pressure at fixed $T$.

These principles will play a crucial role in the following discussion. They must remain valid even for systems that are metastable [8] such as GL. It should be stated once again that we assume that these metastable states are at least in internal equilibrium [8], although they may not be in equilibrium with the surroundings, so that we may describe them by thermodynamic quantities that may very well change in time but the change must be slow.

## III. UNTENABILITY OF THE ENTROPY LOSS CONJECTURE (ELC)

Thermodynamics is a study of a macroscopic system, which is invariably under certain constraints, such as a fixed volume, fixed energy, fixed temperature, absence of certain chemical reaction, proton decay, etc. Such restrictions usually create no problem. All one does is to restrict the allowed phase space to satisfy these constraints, a very common practice. Metastable states require a constraint on time, which necessiates dealing with time-dependent states of the system. At present, there is no consensus on how to deal with non-equilibrium thermodynamics. However, the choice of the experimental observation time $\tau$ allows us to ensure that the system has achieved internal equilibrium (see ELC3), which provides us with a way to describe slowly varying metastable states. We will assume from now on that the system is always in internal equilibrium.

It is true that each replica or sample freezes into one of the basins below $T_g$, but we will not know exactly which one. In particular, if someone exchanged our sample with their own, we will not be able to distinguish as all are characterized by the same set $\xi$. *All of them represent the same macrostate.* Thus, there is a certain amount of uncertainty as to which basin a given sample freezes into. This uncertainty is why we need to use a macrostate description in thermodynamics, whether the system is in equilibrium or out of equilibrium, and which results in the concept of entropy; see Sect. IV for details. Such an argument suggests that the residual entropy should not disappear. It should be stated at this point that ELC is merely a conjecture with no supporting proof or any justifiable calculational or any experimental support in its favor. It is based on the following claim, which totally disregards the above argument:

> ELC: *Since the glass is obtained by confining it in a basin or a metabasin, its configurational entropy is smaller than that of SCL; hence, the configurational entropy discontinuously drops by $\Delta S(T_g) > 0$ at $T_g$.*

It should be stressed that the supercooled liquid at any temperature, at, above or below the transition, cannot know of the choice an experimentalist will make of the duration $\tau$, the observation time; see Sect. V for more details. Thus, whenever its temperature is changed, it will evolve as if it is approaching towards equilibrium that exists at the new temperature in the limited sense in that the crystallization is not allowed. This concept of equilibration in the limited sense will remain implicit in this work as said earlier. Because of the temperature change, this is not a spontaneous evolution. During this approach towards equilibrium for $t > t_f(T, P)$, the system can always be described by applying equilibrium thermodynamics, as the system always remains in internal equilibrium with a well-defined temperature, pressure, etc. that may be different from their equilibrium values. If we now intervene from outside to interrupt the evolution, we end up with GL at $T < T_g$ which is still in internal equilibrium. It has been suggested that ELC is intimately related to ergodicity breaking [2]. The issue of ergodicity is discussed in general terms in Sect. IV C and for glasses in particular in Sect. VII.

For $T \geq T_g$, SCL is always in equilibrium and will turn into a glass by a slight lowering of the temperature below $T_g$, , no matter how small, if the observation time constraint $\tau$ is maintained. Let us now follow the consequences of ELC for a macroscopic system: an abrupt entropy discontinuity *at* the glass transition at $T_g$ (see Fig. 1) with a concomitant discontinuity in the Gibbs free energy $G$ at $T_g$, even though $H$ and the volume $V$ are con-tinuous there; see the blue curves in Fig. 2, which have a discontinuous jump of magnitude $T_\text{g}S_\text{R}$. Afterwards, we will turn our attention to finite systems that are usually studied in simulation in which the sharp discontinuity will be replaced by a smooth variation.

### A. Thermodynamic Inconsistencies due to $\Delta S(T_\text{g}) > 0$

#### 1. Discontinuous Change

(A) The discontinuity in $G$ *violates* the principle of uniqueness of the thermodynamic potential as both the supercooled liquid and the glass are described by the same set of variables: $N, T_\text{g}, P$ and the same history **t**. Thus, both must have the same Gibbs free energy at $T_\text{g}$. Instead, the corresponding thermodynamic potential $G$ is not unique due to the discontinuity.

Loss of ergodicity occurs very often in physics. Common examples are the melting transition, ferromagnetic transition, etc. However, in all such cases, the corresponding thermodynamic potential remains continuous at the transition. In real ferromagnets, there are always magnetic domains, which make these states metastable just as glasses are. No real example is known in Nature where the thermodynamic potential ($G$ in our case) under internal equilibrium has been convincingly demonstrated to show either theoretically or experimentally any discontinuity. Recall hydrodynamics as a prime example, where the continuity of the thermodynamic potential is of paramount importance. The reasons for this important observation will most probably become clear as soon as we realize the resulting violations of thermodynamics and contradiction with experimental facts due to the discontinuity that are discussed below.

(B) Because GL has a different (really much higher) Gibbs free energy at $T_\text{g}$ than SCL, it must represent a different macrostate than SCL at $T_\text{g}$; see ELC2. Their coexistence is impossible as they have different Gibbs free energies, while coexistence requires equal free energies. Thus, the system is either SCL or GL at $T_\text{g}$. In cooling, we have a supercooled liquid at $T_\text{g}$. It now follows from ELC4 and ELC6 that to turn SCL into GL at the same temperature will require a non-zero and finite input of energy from outside either in the form of heat or work. This is most certainly not the case in experiments, where no work or heat input is needed to turn the supercooled liquid into a glass at $T_g$ due to the continuity of $H$ and $V$.

(C) We now consider the issue of how SCL can turn into GL at $T_g$, a distinct macrostate at the same temperature, pressure, and history **t**, but with a higher Gibbs free energy.

  (a) Let us assume that GL is not in equilibrium (with the surroundings) though it may be in internal equilibrium; this is consistent with ELC3. In that case, SCL, which is in equilibrium, cannot leave this state and go to a non-equilibrium state, the glass. Doing this will result in the violation of the second law of thermodynamics according to which a system cannot leave an equilibrium state to go to a nonequilibrium state on its own. This argument provides another but more direct proof the second law violation discussed by Goldstein [1], and is also in accordance with ELC7. It should be noted that the imposition of the observation time $\tau$ cannot do any work or supply heat to the system, while maintaining the temperature at $T_g$.

CONCLUSION 1 *It follows that GL at $T_g$ must be an equilibrium state in order to avoid violation of the second law. This violation occurs regardless of whether $G$ is continuous or discontinuous at $T_g$.*

This conclusion contradicts ELC3, according to which GL at $T_g$ has been assumed to be in internal equilibrium, but not in equilibrium.

(C) We now follow how SCL can turn into GL at $T_g$, if GL is in equilibrium.

  (b) An equilibrium GL will contradict ELC3. Because of the continuity of the enthalpy and volume at $T_g$, there can be no exchange of heat and no production of work in the transition; see (B). Thus, there is no way for SCL to raise its Gibbs free energy and turn into GL or vice versa.

We can now draw the following

CONCLUSION 2 *It follows from (C(a)) and (C(b)) that in a cooling experiment, which is what we consider here, there is no way to convert SCL into GL at $T_g$.*

(D) A discontinuity $\Delta_\mathrm{g} G = -T_\mathrm{g} S_\mathrm{R}$ in $G$ also makes $\gamma \equiv G/T$ discontinuous with a discontinuity $\Delta_\mathrm{g}\gamma = -S_\mathrm{R}$. As a consequence of the discontinuities, the derivatives $(\partial G/\partial T)_P$ and $(\partial \gamma/\partial T)_P$ at $T_\mathrm{g}$ with respect to $T$ become infinite as can be seen from the following. Consider a discontinuous quantity $X$ $(= G$ or $\gamma)$ at $T_\mathrm{g}$ and the limit

$$\underset{\Delta T \to 0}{Lim} \frac{X(T_\mathrm{g} + \Delta T/2) - X(T_\mathrm{g} - \Delta T/2)}{\Delta T} = \underset{\Delta T \to 0}{Lim} \frac{\Delta_\mathrm{g} X}{\Delta T} + \frac{1}{2}[X'(T_\mathrm{g+}) + X'(T_\mathrm{g-})] \to -\infty,$$

where $X'(T_\mathrm{g+})$ and $X'(T_\mathrm{g-})$ are the right and left derivatives at the discontinuity. From the thermodynamic relations [30]:

$$S = -(\partial G/\partial T)_P, \quad H = -T^2(\partial \gamma/\partial T)_P, \qquad (7)$$

we conclude that a discontinuous $G$ results in the entropy and enthalpy to become *infinitely large* at the glass transition. However, it is a common knowledge that both quantities remain bounded at the glass transition.

2. *Continuous Variation and Instability*

The above inconsistencies with thermodynamics or experiments occur because of the discontinuity. It is well known that similar discontinuities in entropy, volume etc. occur only when we take the thermodynamic limit $N \to \infty$ [24, see p. 206]. In all these cases, the same quantities remain perfectly continuous for a *finite* system such as in any simulation (though experimentally, one may not be able to distinguish them from a discontinuous quantity due to experimental errors, unless one considers a small enough system). Then, it may appear that the above inconsistencies will not emerge if we do not take a thermodynamic limit, which certainly will be gratifying as all real systems are finite: $N$ is large but not infinitely large. This is indeed what is seen in a recent model computation of a finite system of selenium by Mauro et al [14, and references therein of the previous work of this group] using several cooling rates. They consider a system of 64 particles and compute the continuous drop in the entropy. Their model calculation of the finite system identifies the melting temperature at $T_\mathrm{m} = 490$ K, and they consider $r = 10^{12}$ K/s, $10^6$ K/s, 1 K/s, $10^{-6}$ K/s, and $10^{-12}$ K/s. For the fastest cooling rate, the glass transition occurs almost at the melting temperature, while for the slowest cooling rate, it occurs at $\simeq 270$ K. Thus, the maximum entropy drop is almost equal to the melting entropy. One is struck by the range of temperature over which the drop occurs. It is $\simeq 200$ K for all of the cooling rates that have been reported, even though the cooling rates have changed by a factor of $10^{24}$, an enormous factor. Thus, there is no hint that the enormous width of $\simeq 200$ K at all would shrink to zero as the cooling rate $r \to \infty$ (instantaneous cooling). Moreover, as the glass transition cannot be below the melting temperature (it appears to be almost at $T_\text{m}$ for $r = 10^{12}$ K/s, one is forced to conclude that this cooling rate is almost equivalent to $r \to \infty$. Thus, it is hard not to conclude from their results that the enormous and almost constant width and the continuous fall in the entropy is a *finite size effect*, as we have shown in Figs. 1 and 2. The continuous drop can only become discontinuous as the thermodynamic limit is taken. Unfortunately Mauro et al [14] have not carried out any finite size analysis, so it is not possible to verify if what they see is a finite size effect, or a finite cooling rate effect. We now investigate the consequences of such continuous variations in the entropy and the Gibbs free energy, regardless of whether they are due to finite sizes or finite cooling rates.

(E) A careful analysis of this issue points to a few other problems with ELC. To see this, let us inquire how such a discontinuity can emerge in a thermodynamic limit. In other words, we ask: How such a discontinuous Gibbs free energy can emerge from a continuous Gibbs free energy of a finite system as its size becomes infinitely large. We should remark that the Gibbs free energy is a concave function of its arguments, as the blue curves are in the figure; see (6). Because of this, the entropy and specific heat $C_P$ remain not only positive but also finite. The free energy must always curve downwards for the system to remain stable. Now, the two blue branches will deviate from its form for a finite system. We have shown one such possibility by the red curve for a finite system. It contains three different pieces: the two solid pieces in Fig. 2 on either side of the dot-dashed piece in the middle containing an inflection point A where the curvature changes its sign. The solid pieces are concave as required by stability. But it is also clear that any attempt to connect the two solid pieces by the dot-dashed piece near $T_\text{g}$ must result in the middle piece with the following properties:

   (a) The middle piece cannot remain concave everywhere. It must have a convex piece at higher temperatures, which will then result in a region of *negative heat capacity* as shown by the arrow, and which will make the system *unstable* there.

   (b) At the inflection point A (there is another inflection point at a higher temperature,

which is not important in our discussion), the slope becomes very large and negative, so that the entropy $S$ will become extremely large. This large negative slope at A is what will ensure a discontinuity in the Gibbs free energy in the limit $N \to \infty$. However, for the system to possess such a huge amount of entropy at low temperatures is certainly unthinkable and contrary to all experimental facts. Similarly, the enthalpy will become very large at the inflection point of $\gamma$. In the limit $N \to \infty$, there result diverging entropy and enthalpy in accordance with (7).

### B. Continuity of $G$ and $S$ from the continuity of $H$ and $V$

The above discussion assumes a discontinuity in the Gibbs free energy or the entropy. One may claim that the discontinuity in $G$ or $S$ is real, not withstanding the above pathologies.

(F) We now argue that a discontinuity in $G$ is inconsistent with a continuous $H$ as a function of $T$, including $T_\mathrm{g}$. The latter is an experimentally observed fact for glasses. The continuity of $G$ follows from the thermodynamic identity

$$G(T) = G(T_0) + T \int_{1/T_0}^{1/T} H(T) d(1/T), \tag{8}$$

which is obtained from the second equation in (7), and clearly shows that the Gibbs free energy must also be a *continuous function* of $T$, in direct contradiction to ELC, even when the history of the process is accounted for. Incidentally, (8) also constitutes a proof of the principle of uniqueness of $G$ for metastable supercooled liquids. To make ELC comply with the continuity requirement will require a continuous $G$ in the glass transition region with the loss of entropy occurring in a smooth manner (without a discontinuity) in this temperature region. Such a scenario will then result in $G$ to be replaced by a smooth curve such as the red curve in Fig. 2. Thus, the problem with negative specific heat and abnormally large entropy will persist.

(G) If $G$ is continuous as a function of $T$, then the entropy cannot be discontinuous, which again contradicts ELC. We now give a separate argument in favor of a continuous entropy. For this, we will use the fact that the volume $V$ is continuous across the glass transition, and use the thermodynamic identity

$$(\partial S/\partial P)_T = -(\partial V/\partial T)_P. \tag{9}$$

The form of $V$ found in experiments makes the right-hand side finite and negative. We now argue that this implies that $S$ cannot be discontinuous. We will apply the above identity at $T = T_g$ and the experimentally chosen $P$. To determine the pressure derivative, we consider two different pressures $P_1 = P + \Delta P/2$ and $P_2 = P - \Delta P/2$, $\Delta P > 0$, each with its glass transition temperature $T_{g1}$ and $T_{g2}$, respectively. It is normally the case that $T_{g1} > T_g > T_{g2}$. The derivative $(\partial S/\partial P)_{T_g}$ is evaluated by the limit of $(\Delta S/\Delta P)_{T_g}$ as $\Delta P \to 0$. For $P_1$, $T_g$ lies on the glass side, while for $P_2$, $T_g$ lies on the supercooled liquid side. Thus, $\Delta S \equiv S_1(\text{glass}) - S_2(\text{liquid}) \to -S_R$ as $\Delta P \to 0$. Consequently,

$$(\partial S/\partial P)_{T_g} \to -\infty, \tag{10}$$

which makes the expansion coefficient infinitely large. This is most certainly not the case in experiments and clearly shows that ELC is *inconsistent with the continuity of volume* at the glass transition. Replacing the discontinuity in $S$ by a continuous patch around $T_g(P)$ will still give rise to a very large negative $(\partial S/\partial P)_{T_g}$, and a large expansion coefficient, which does not seem to be realistic near a glass transition.

(H) The discontinuity in $S(T, P_0)$ at $T_g$ as we vary $T$ for a fixed pressure, which we now denote by $P_0$ for reason that will become clear in a moment, denotes a discontinuity at $T_g, P_0$ in the $TP$ plane. Thus, this discontinuity will also be encountered if we fix the temperature at $T_g$ and vary the pressure. We now argue that this discontinuity is inconsistent with the continuity of the derivative

$$V'_g(P) = (\partial V/\partial T)_P|_{T=T_g(P_0)},$$

which is the derivative calculated at a fixed temperature $T = T_g(P_0)$ but for different pressures $P$. This requires conducting different experiments at different pressures including the original pressure $P_0$. Experiments show that $V'_g(P)$ is a continuous function of $P$. Integrating the identity $(\partial S/\partial P)_{T_g} = -V'_g(P)$ from $P' < P_0$ to $P > P_0$, we find that

$$S(T_g, P) = S(T_g, P') - \int_{P'}^{P} V'_g(P) dP,$$

which shows that the entropy must remain continuous across $P_0$ due to the continuity of $V'_g$.

## C. Stability violation from a concave $S$

(I) Now that we have demonstrated that both $G$ and $S$ are continuous, let us determine the conditions that must be satisfied by $H$ and $S$ to ensure stability. The latter requires the curvature, the second derivative, of $G$ with respect to $T$ to be negative. Now, experimentally, it is observed that $H$ has a positive curvature near $T_g$. Thus, from the identity

$$\left(\frac{\partial^2 G}{\partial T^2}\right)_P = \left(\frac{\partial^2 H}{\partial T^2}\right)_P - T\left(\frac{\partial^2 S}{\partial T^2}\right)_P, \tag{11}$$

we conclude that the curvature of $S$ should not only be positive, but positive enough to make the the curvature of $G$ negative. In particular, $S$ cannot have a negative curvature near $T_g$, which will make the curvature of $G$ positive and lead to instability. Now, a possible continuous approximation of the discontinuity in $S$ is shown by red dashed portion in Fig. 1 in the neighborhood of the glass transition. Any such continuous $S$ will lead to a negative curvature near the supercooled liquid region, leading to the stability violation, just as was the situation seen in points (D) and (F).

At this point, we turn to the calculation of the residual entropy by Mauro et al [14] for five selenium glasses using their landscape model. The calculation is performed for a finite system with a constant vibrational frequency for all basins, and the resulting residual entropy is found to vary *continuously* as we have suggested due to the finite size effects. The discontinuity has disappeared. To use (11), we need the entropy, which requires adding the vibrational entropy to the calculated residual entropy. Unfortunately, Mauro et al. do not give the vibrational entropy or the entropy. Thus, we are forced to estimate the vibrational entropy by considering GL as a solid in which particles are vibrating in a harmonic potential. As the authors only consider topologically identical basins [14, See Ref. 52 for details], the vibrational frequency in the harmonic approximation remains independent of the pressure and temperature. We then find that the vibrational entropy in this approximation is given by

$$S_{\text{vib}}(T) = 3N \ln T + \cdots, \tag{12}$$

where we are not showing terms that are constant; see [8, Eq. (65.8)]. Thus,
$$\left(\frac{\partial^2 S_{\text{vib}}}{\partial T^2}\right)_V = -\frac{3N}{T^2} < 0.$$

The curvature at constant $P$ will not change for the reason mentioned above. The residual (configurational) entropy obtained by Mauro et al [14] has a negative curvature below the glass transition, which is enhanced by the negative curvature due to the vibrational entropy. Thus, their model calculation for a finite system leads to a negative curvature of the entropy $S$, leading to a negative specific heat. In other words, their calculation violates stability, as we have concluded on general grounds if ELC is accepted.

To be sure of the above conclusion, we need to know more about any temperature dependence of the term not shown in (12). Again, Mauro et al provide no help in this direction. So, we turn to the numerical work by Mossa et al [31] in which a fragile glass former is studied using the potential energy landscape picture [5]. As the distribution of observed inherent structure is found be Gaussian, and the basins have almost same shape, just as Mauro et al discover in the enthalpy landscape, we can use the results in ([31]) without much reservation. The missing term above is the part of the second term in the equation (11) in ([31]), and is found to be linear in $T$, as seen from Fig. 9(a) there. Thus, this correction in the harmonic approximation does not affect the above curvature, so the conclusion remains unchanged. These authors also consider anharmonic correction to the vibrational entropy; see their equation (33) and Fig. 10. As the curvature of this contribution is found to be negative, this correction also does not change our above conclusion that the smooth variation of the entropy for the finite system violates stability.

The results by Mauro et al is merely a reflection of our general proof of the violation of stability in a finite system if the calculation is based on ELC. Thus, even if one is not sure of the our discussion of the curvature of the calculated entropy by Mauro et al, one must remember that the violation of stability must occur in any finite system calculation, which is based on ELC. This should be contrasted with the behavior of the true entropy of a finite system, which will never show any loss of stability.

The above conclusions are based on repeated use of thermodynamic relations and the principles laid down in Sect. II. The last three points are solely based on experimental facts regarding the continuity of $H$ and $V$. From these points, the following conclusion is inescapable:

> CONCLUSION 3 *The continuity of $G$ for a finite system is not sufficient by itself to make it thermodynamically consistent. It must also remain concave. The latter can only happen if the original discontinuity in $G$ or $S$ for a macroscopic system were to vanish.*

The following point should again be emphasized. Continuous $G$ or $S$ above have been associated with finite systems. One can argue that this continuity is also a property of a macroscopic system for most values of $r$. In that case, we must inquire if $G$ remains concave or not. For stability, it must remain concave, or $S$ must not be concave. Thus, the CONCLUSION 3 is valid for continuous $G$, regardless of how it arises.

### D. ELC must be rejected

One can continue to find other inconsistencies that ELC creates with classical thermodynamics. For example, classical thermodynamics treats $S$ as a differentiable function of the energy $E$. Now, from the continuity of $H$ and $V$, one concludes that $E$ is also continuous across the transition. If $S$ were a continuous function of $T$, then one can express $E(T)$ as $E(S)$, which can be inverted to yield to give $S(E)$. This is not possible if ELC is accepted. We believe that we have already constructed a long list of violations, either thermodynamic or experimental, to make our case. Thus, it is safe to conclude that ELC, either with a discontinuity or with a smooth patch, is untenable and must be rejected if we want to exploit thermodynamics and remain consistent with experimentally observed and verified facts. Of course, one can also claim that one must develop a new kind of thermodynamics that applies to glasses just to salvage ELC. In that case, we have nothing to say that would be useful, as we do not know the form of the new thermodynamics. But if the customary thermodynamics is adopted, then the recent calculation by Mauro et al [14] discussed above already shows that it leads to stability violation. Accordingly, we are going to take a more conservative view that there is nothing wrong with thermodynamics. In that case, we need to develop a formalism to explain how the entropy can remain continuous even when the glass is confined to a basin. Such a formalism should also provide answers for Q1 and Q2.

## IV. GENERAL FORMALISM

### A. Thermodynamic Averages and the Ensemble Interpretation

Before proceeding further, we need to set the stage for our arguments. Our approach is based on concepts that have been developed and tested in equilibrium statistical mechanics [8, 9, 23, 24], which we now extend below to time-dependent systems; see also Sethna [10], Jäckle [17], Palmer [18] and Rice [32]. Our starting point is the principle of reproducibility for any system, whether in equilibrium or time-dependent; see Sect. II and the FUNDAMENTAL AXIOM. It is this viewpoint that we adopt in this work. This answers our first question Q1 about any average, which as we will see below contains entropy, an average quantity; see (15) and (19). We formalize the averaging over *independent* samples or replicas as follows. Let us consider an $N$-particle system $\Sigma$, which has $W$ distinct microstates $j$ of energy $E_j$, $j = 1, 2, \cdots, W$. *All samples have the same history* $\mathbf{t}$. At each point in the history $\mathbf{t}$, each sample is in one of the $W$ microstates (unless the samples have been specially prepared not to be in all of the microstates $W$). Let $N_j(\mathbf{t})$ denote the number of samples in the $j$th microstate at time $t$, so the the probability

$$p_j(\mathbf{t}) \equiv N_j(\mathbf{t})/N_\text{S}, \quad \sum_{j=1}^{W} p_j(\mathbf{t}) \equiv 1, \qquad (13)$$

denotes the probability of the $j$th microstate; the probability in general depends on time. Note that the microstates and their energies do not depend on time or other external field variables like the temperature, pressure, etc. but their probabilities in general do for time dependent systems. As is well known, the above probabilities require the formal limit $N_\text{S} \to \infty$, which is going to be implicit in the following. In reality, the value of $N_\text{S}$ does not have to be large for obtaining a reliable result when we are dealing with systems in equilibrium. All macrostates contributing to the equilibrium are sharply peaked at the equilibrium values. For each of these macrostates, the microstates are equally probable. Thus, the probability of finding a sample in one of these microstates will be large. Hence, even a small value of $N_\text{S}$ will be sufficient to give us a highly reliable result. No such guarantee is available when the system is not in equilibrium.

The average $\overline{X}$ of $X$ is given by

$$\overline{X}(\mathbf{t}) \equiv \sum_{j=1}^{W} p_j(\mathbf{t}) X_j, \qquad (14)$$

where $X_j$ is the value of $X$ in the $j$th microstate; the value $X_j$ also does not depend on external field variables if $X$ does not. The above sample average also follows immediately from the principle of additivity. One considers a very large system $\Sigma_0$ of $N_0 \equiv NN_{\rm S}$ particles and imagines dividing the large system into a large number $N_{\rm S}$ of macroscopically large parts of equal size $N$, representing various microstates of the system $\Sigma$. As the parts are macroscopically large, they will act almost independently, which is a prerequisite; see above. How well this condition is satisfied depends on how large the parts are. In principle, they can be made arbitrary large to ensure their *complete independence*. At the same point in history $\mathbf{t}$, these parts will be in microstates $j$ of $\Sigma$ with probabilities $p_j(\mathbf{t})$. One can think of the $N_{\rm S}$ parts as representing the same measurement that has been repeated $N_{\rm S}$ times on samples prepared under identical macroscopic conditions. We simplify our notation and use $t$ to denote the entire history $\mathbf{t}$ from now onwards. This should cause no confusion.

In a measurement such as a calorimetric measurement, each part will contribute in the average (14). For example, the amount of heat $\Delta Q$ given to the system at constant pressure over a duration $\Delta t$ is given by the difference of two instantaneous enthalpies

$$\Delta Q = \overline{H}(t + \Delta t) - \overline{H}(t).$$

(It is a common practice to denote thermodynamic averages such as of $H, G$ etc. without the bar in the literature, so we will now follow this practice.) A measurement of enthalpy at some given instant requires an average carried out at that moment over various parts or samples. Of course, we are assuming that the measurements can be performed instantaneously, an assumption commonly made in physical sciences.

For an event which occurs with probability $p$, $\ln p$ denotes what Gibbs [26] calls the *index of probability*; Shannon [27] identifies $-\ln p$ as the amount of *uncertainty* (not to be confused with Heisenberg uncertainty in quantum mechanics); the index or the amount of uncertainty is a property of the system. At the same time, each represents an additive quantity (like the entropy) for independent events (parts or samples), and can be used to define the entropy $S(t)$ of the system, which at any given instant $t$ is nothing but the average of the uncertainty

$-\overline{\ln p}$ :
$$S(t) \equiv -\overline{\ln p} \equiv -\sum_j p_j(t) \ln p_j(t). \tag{15}$$

It represents the average uncertainty over all the parts or samples as follows from our FUNDAMENTAL AXIOM. This formulation of entropy is valid regardless of what thermodynamic variables $\xi$ are used in the thermodynamic description of the system. Its value depends on the time-dependence of the probabilities, and it can be easily shown that it cannot decrease with time; see Tolman [9, Sect. 106, where Boltzmann's $H = -S$ is considered], and Rice [32, Ch. 17]. If we use all extensive quantities (besides $t$) such as the energy $E$, volume $V$, number of particles $N$, etc. then we are considering what is traditionally called an isolated system. For such a system, we can state the second law of thermodynamics in terms of this entropy, whose negative plays the role of the thermodynamic potential for these variables: $S$ is a *non-decreasing* function of time. It approaches its maximum allowed value only when the system is in *equilibrium*, which occurs when

$$p_j(t) \stackrel{t \to \infty}{\Rightarrow} 1/W \quad \text{for each microstate } j. \tag{16}$$

In that case only, the entropy is given by the Boltzmann relation

$$S(t) \stackrel{t \to \infty}{\Rightarrow} \ln W. \tag{17}$$

We wish to emphasize that, according to our formulation, a system is in equilibrium when its ensemble entropy is at its maximum; see [7]. According to the principle of reproducibility, this means that a thermodynamic system also has its entropy at its maximum (in the average sense). The second law then states that this entropy can never decrease [7]. If we replace any of (but not all of) the extensive quantities by their conjugate field variables such as the temperature $T$, pressure $P$, chemical potential $\mu$, etc., then we must state the second law in terms of the corresponding thermodynamic potentials, which are *non-increasing* functions of time. We refer the reader to the textbook by Landau and Lifshitz [8, Sect. 15] or by Huang [24, Sect. 1.6] for a general discussion of the above statement. For the description requiring $N, T, P$, and $t$, with which we are interested here, we give a simple proof for $G$ in Sect. VI, which we will apply to glasses.

The approach discussed above is the conventional ensemble approach due to Gibbs [26], and is applicable to time-dependent systems as well. Because of this, we will call $p_j$ the instantaneous *ensemble* probability (in the limit $N_\mathrm{S} \to \infty$), and the thermodynamic average $\overline{X}(t)$ as the instantaneous *ensemble* average. If $p_j(t)$ are known, we can use (15) to find the time-dependent entropy or use (14) to find other average quantities. Thus, the main goal in non-equilibrium thermodynamics is to determine $p_j(t)$.

The identification of entropy in (15) with the Gibbs formulation of entropy is a time-honored practice since the days of Gibbs [26, see, in particular chapters 11 and 12, where time-dependence is discussed], and has been discussed by Tolman [9, Ch. 13, and in particular pp. 538-539], Jaynes [25], Rice [32], to name a few. There is no restriction on $p_j(t)$; in particular, they do not have to be given by (16). Thus, our formulation is equally suited to study non-equilibrium situations; see also Sethna [10, Sect. 5.3.1]. The idea of ergodicity has never entered in its formulation. It merely follows from the observation that the index of probability is an *additive* quantity for independent replicas (see FUNDAMENTAL AXIOM) and that the entropy is merely its average value (with a negative sign). Tolman takes great care in establishing that this formulation of the entropy satisfies the second law [9, Sect. 130]. Tolman also shows that the Boltzmann definition of entropy is a special case of the general formulation due to Gibbs [9, see the derivation of eq. (131.2)], just as we have argued above in regards to (17). Thus, we find that ELC9 is not really a proper reflection of the generality of the Gibbs entropy formulation. The identification of the entropy with the negative of the Boltzmann $H$-function [9, see p. 561], the latter describing a non-equilibrium state, should leave no doubt in anyone's mind that the Gibbs formulation of the entropy can be applied everywhere. Nevertheless, we should point out that not all subscribe to this viewpoint of ours about the Gibbs formulation of entropy, see for example [2, 14], because they insist that the Gibbs entropy is a constant of motion. This constancy follows immediately from the application of Liouville's theorem in classical mechanics [8, 9, 24] valid for a system described by a Hamiltonian; see Sect. IV D also, where we will see that even our interpretation of the entropy is consistent with this theorem.

### B. Role of Dynamics and the Temporal Interpretation

We now turn our attention to answer our second question Q2. In our ensemble approach involving independent samples, parts or replicas, each sample is independent of and remains uninfluenced by all other samples. The approach does not even require any knowledge of the actual dynamics governing the system. In particular, the concept of entropy as an average uncertainty in the ensemble approach is ambivalent to the presence or absence of any dynamics in the system. Of course, in the latter case, the entropy will not change with time. We can get a better appreciation of this irrelevance by considering a system of $N$ *non-interacting* Ising spins, for which $W = 2^N$ denotes the number of distinct microstates, each microstate having the same *a priori* probability $p_j = 1/W$. Whether there is any dynamics specified or not is irrelevant in determining the entropy. All of us will agree that the entropy is $S = N \ln 2$. All of us will also agree that it is the maximum entropy. This entropy will be the same even if there were no dynamics changing the spins, a situation that also occurs at absolute zero as far as the classical dynamics is considered. This situation can also be identified with $t_\text{s} \to \infty$. Accordingly, every sample will remain in its microstate forever, just like a system which remains confined to a basin due to kinetic freezing ($t_\text{s} \gg \delta$, which is almost identical to $t_\text{s} \to \infty$). However, as we have said above (the FUNDAMENTAL AXIOM) and as Tolman has emphasized, the entropy is an *average* quantity obtained by an average over *all* samples or microstates. We cannot just consider one particular sample. This is equivalent to saying that the entropy is determined by the macrostate, which represents a collection of microstates, each with certain *a priori* probability. The entropy has a contribution from all of these microstates. It is not the property of a single microstate. This observation is unaffected by whether microstates have any dynamics to change them or not in time. This thus answers Q2. This is consistent with the entropy $S = N \ln 2$ for noninteracting Ising spins.

For systems that are expected to be in equilibrium, it does not matter when repeated measurements are made. However, when we deal with measurements on a glassy system, which continues to change slowly with time, it becomes crucial to ensure that the repeated measurements are made simultaneously at the same time; otherwise, different measurements will be for systems that cannot be called identically prepared. This makes our current discussion very different from what is conventionally done in equilibrium thermodynamics, where one has a tendency to treat identically prepared samples representing the same system that may have been prepared at different times. This equivalence is then used to justify that the sample average is not different from the average over time. The *temporal* average, however, can only be useful if some dynamics is provided for the system under which microstates evolve in time. Moreover, as said above, the temporal average is not unique; there is an ar-bitrariness in it due to the number of initial samples used in its definition, as we will discuss below. Let us, for the moment, assume observing the evolution of $\Sigma$ in time, which is known to be in a particular microstate $j_0$ at the initial time $t = 0$. The instantaneous microstate into which the initial microstate $j_0$ is evolved is recorded at discrete times $t_k = k\delta$, where $\delta$ is the average time required for a microstate to change to another microstate and $k \geq 1$ is an integer. Then, after a long period of time, these microstates (at different times) can be thought of as representing different parts of $\Sigma_0$ at some particular moment. Such an interpretation, which is common in equilibrium statistical mechanics, allows us to introduce a *temporal* average and entropy as follows. Let $N_j(t_k)$ denote the number of times the initial microstate $j_0$ evolves into the microstate $j$ during the time interval $(0, t_k)$, which is used to determine the temporal probability

$$\widehat{p}_j(t_k) \equiv N_j(t_k)/N(t_k), \ k \geq 1, \tag{18}$$

where $N(t_k) = k$ is the total number of microstates at time $t_k$. These probabilities, which we call *temporal* probabilities, in general will depend on $j_0$, and can be used to introduce the *temporal* average $\widehat{X}$ of $X$:

$$\widehat{X}(t_k) \equiv \sum_j \widehat{p}_j(t_k) X_j, \ k \geq 1;$$

here $X_j$ is the value of $X$ in the $j$th microstate as in (14). Thus, the temporal entropy is defined as

$$\widehat{S}(t_k) \equiv -\overline{\ln \widehat{p}} \equiv -\sum_j \widehat{p}_j(t_k) \ln \widehat{p}_j(t_k), \quad k \geq 1, \tag{19}$$

which formally looks identical to the form in (15), except that $p_j$ is replaced by $\widehat{p}_j$. However, the most important difference is that as $\widehat{p}_j$ is defined over a duration of time, its usefulness is only when a dynamics exists so that microstates can change, while $p_j$, being an instantaneous quantity, is oblivious to the dynamics.

The choice of the initial microstate $j_0$ is not unique. So, one can start with any microstate $j'_0$ as the initial microstate. Similarly, we can follow the temporal evolution of not one but many distinct microstates selected simultaneously at $t = 0$, define $N_j(t_k)$ to denote the number of times this collection of microstate evolves into the microstate $j$ during the time interval $(0, t_k)$ and $N(t_k)$ the total number of microstates at time $t_k$. Using these quantities in (18) will give yet another value of the probability $\widehat{p}_j(t_k)$, which will depend on the collection of initial microstates. These probabilities will give a different average $\widehat{X}(t_k)$ and entropy $\widehat{S}(t_k)$. Thus, we can introduce many different temporal quantities depending not only on the number of initial microstates, but also which ones are the initial microstates. These quantities will also depend on the history **t**.

> REMARK: *We emphasize that the ensemble definition (15) of the entropy or the average (14) only requires the probabilities of microstates, but does not require a non-zero probability of a transition from one microstate to another. This is important to remember when we apply this definition to glass forms which are considered to be frozen in that they do not jump from one glass form to another during a short period of observation ($\delta >> \tau$). On the other hand, for the temporal definition to be useful requires not only for the microstates to change with time due to a dynamics, but also requires the time $t = t_k >> \delta$ ($k >> 1$). It should also be stressed that there is no unique temporal average or entropy as one can take any of the microstates as the initial microstates in any number.*

## C. Ergodicity, Its Breaking and Restoration

It is evident that both, ensemble and temporal, definitions of the average or entropy require the same average over microstates. The only difference is in the form of the probability used for the microstates. Any difference between the two microstate probabilities results in a difference between the two averages. From the discussion above, it appears that the ensemble approach appears more fundamental than the temporal approach, mainly because of the additivity principle. The other reason is that most measurements last a short period of time. The temporal average over an extended time period has nothing to do with information obtained in measurements that may take a fraction of a second or so. In contrast, the ensemble average provides an instantaneous average and thus bypasses the above objection of the finite measurement time. The temporal average over a short period will only make sense if it remains equal to the ensemble average; see the principle of reproducibility in Sect. II. This can only happen if the system is in equilibrium to begin with. For metastable or time-dependent states, its usefulness is quite questionable. To be able to carry out a temporal average, we need to prepare the system in a particular microstate $j_0$ by carrying out a microstate measurement on the system that was introduced above in Sect. IV A. It may then take quite some time, much longer than the experimental time, before the temporal average can come close to the ensemble average. As the dynamics becomes too slow in the glassy state, this time may become astronomically large. One such situation is discussed later in Sect. VII. Thus, temporal average may not be desirable in general. We refer to Becker [23, p. 116], Tolman [9, p. 69], and Jaynes [25, first paragraph, p. 106] for additional information on this point.

Despite the lack of any real superiority of the temporal average, many people consider it to be of primary importance. Then, they need to justify its equivalence with the ensemble average, which then leads to the concept of ergodicity. A system is said to be *ergodic* if the *limiting* ensemble average $\overline{X}$ as $N'$ or $N_\text{S} \to \infty$ is equal to the limiting temporal average $\widehat{X}$ (with any number of initial microstates, even though one customarily considers a single microstate) as $k \to \infty$, provided $t_\text{s}$ remains finite. They are usually believed to be the same when the system is in equilibrium. This requires that the limit of $\widehat{p}_j(t_k)$ is equal to the limit in (16). In most cases of interest, $t_\text{s}$ is finite so that the system comes to equilibrium in a finite amount of time. Let us introduce a particular value $k_\text{eq}$ by

$$k_\text{eq} \equiv t_\text{s}/\delta. \tag{20}$$

One only needs to consider

$$k \gtrapprox k_\text{eq} < \infty, \tag{21}$$

instead of the limit $k \to \infty$, to ensure that the temporal average equals the ensemble average in a *finite* duration; as we will see in Sect. IV D, we are considering stochastic dynamics so that there is no recurrence. In such a situation, one can use either of the two averages to obtain proper thermodynamics.

It should be evident that the concept of ergodicity has a meaning only in the infinite time limit. But this limit may not commute with $t_\text{s} \to \infty$. Above, we had considered $t \to \infty$ via $k \to \infty$, while $t_\text{s}$ was kept finite and fixed. The situation will be drastically different if $k_\text{eq} \to \infty$; this occurs if $t_\text{s} \to \infty$. The same situation also occurs when $\delta = t_\text{s} > \tau$, a situation which arises for glasses. In both cases, the temporal limit will not reproduce the ensemble limit, and the temporal evolution will not converge to the equilibrium situation. If it happens that the two limits are different ($k \to \infty$), then the system is said to be *non-ergodic*.

It usually is the case that the loss of ergodicity occurs because the sets of microstates in the two averages happen to be different. It appears from our discussion in Sect. IV A that the two averages would be identical in equilibrium if they contained the same set of microstates. This gives us the clue to restore ergodicity for equilibrium states when it is lost. In the latter case, there is a standard way by introducing what is called a symmetry-breaking field to *restore* ergodicity. The role of the new field is to restrict the set of microstates so that they are the same in both averages. This observation of somehow ensuring the sets of microstates to be the same will be central later when we revisit ergodicity in glasses in Sect. VII. Thus, once ergodicity has been restored by a proper choice of microstates (an example of this will be seen in Sect. IV D when we elect to have $w = W$), we can follow either of the two averages to investigate its thermodynamics. However, even in this case, the temporal average will still require a very long time to satisfy ergodicity and will not be suitable for experiments that last only a short duration of time.

It is important to mention that Gibbs himself did not find the concept of ergodicity relevant to the foundation of statistical mechanics, as it finds no place in his monumental work [26]. Our view is to ensure that the ensemble average is equal to the experimental values; see principle of reproducibility in Sect. II. Whether the ensemble average is equal to the time average is of no use to an experimentalist since most experiments do not take that long to perform and, as has been known, see for example Jaynes [25, second paragraph, p. 106], and Goldstein [1] who visits this argument again, that the time required to sample all possible microstates is much longer than the age of the universe.

It should be clear from the discussion in the previous section that the issue of ergodicity is relevant not only when some dynamics is available for microstates to change, but also when we are dealing with equilibrium states because of the infinite time limit needed for the temporal average. To extend temporal average to non-equilibrium or metastable states (finite but long $t_{\text{s}}$) over a *finite* period of time

$$k < k_{\text{eq}} < \infty,$$

and to compare with the instantaneous ensemble average has no relevance for the issue of ergodicity ($k \to \infty$). Despite this, there is a strong tendency in the glassy field to considers the glass transition to represent loss of ergodicity; see ELC3 [2, 15]. Therefore, the issue of ergodicity for glassy states requires careful analysis, which we defer to Sect. VII. But we consider the temporal entropy in its general form in Sect. IV D and show how unsatisfactory the temporal entropy can be.

It should also be stressed that the temporal average over infinite time is most certainly inappropriate for glasses, which over a long period of time will relax to their equilibrium supercooled liquid states or to the ideal glass if left to themselves. Thus, the temporal average carried out over an infinitely long period will describe the equilibrated supercooled liquid or the ideal glass and not the glass, obtained under an experimental time constraint $\tau$, unless appropriate theoretical restrictions are imposed as we will discuss later in Sect. VII. The ensemble average does not suffer from this problem, which therefore becomes the choice average to consider for studying glasses. In any case, the issue of ergodicity does not seem to be very relevant for laboratory glasses, more so than it is not relevant for equilibrium states [9, Sect. 25]; we will come back to this issue later in Sect. VII. We find that there is no loss of ergodicity in our approach if proper care is exercised in identifying the microstates. Thus, the only important issue is to ensure that our interpretation is consistent with all the principles of thermodynamics as the latter has been tested over and over again and found to be always valid.

### D. Microstate Measurement, Probability Reduction, and Temporal Entropy

As long as we are content with a macrostate description of a system, a description that is incomplete for the system, we have no need to know which microstate the system is at a given instant. To identify a particular microstate requires complete information about the system, which is ordinarily unfeasible. In order to identify any particular microstate, we need to perform a very special kind of "measurement," which we will call a *microstate measurement*, that provides us with the complete information about the system in its current microstate $j_0$. Such a situation might prove relevant for a system kinetically frozen in a *particular* basin. In that case, we are mainly interested in the dynamics of the system as it moves from this basin to another. For glasses, $N_\text{b}$ will play the role of $W$. Because of this relevance, it is necessary to investigate such a special preparation. For the Ising model, this requires *determining* the orientations of each of the $N$ spins. After the microstate measurement, we know with certainty which microstate the system is in. Accordingly, its probability changes *discontinuously* from $p_0 = 1/W$ before the measurement to $p_0 = 1$ immediately after the measurement. The effect of the microstate measurement is to also reduce the probabilities of all other microstates $j' \neq j_0$ to $p_{j'} = 0$. Thus,

$$p_j = \delta_{jj_0}, \tag{22}$$

where $\delta$ is the Kronecker delta, immediately after the measurement. We will speak of the *probability reduction* to indicate this change in the probability brought about by the microstate measurement in this work. The entropy also vanishes in an abrupt fashion immediately after the "measurement" from the initial value of $\ln W$ in accordance with the complete certainty about the system. The above idea of probability reduction is not a novel idea and follows from common sense. It is not surprising that it is widely accepted in the field. For example, Tolman assumes this probability reduction when he discusses this kind of measurement to identify a particular microstate [9, see the discussion immediately below Eq. (104.14)]. The same probability reduction is also invoked by Mauro et al [14] after a microstate has been identified by a measurement.

At present, we will use this observation to discuss equilibrium or lack of it and the role of ergodicity in glasses. The existence of ergodicity is usually equated with equilibrium. Its absence may or may not imply a lack of equilibrium. For glasses, which is our interest here, its absence is also used to imply irreversibility; see [2, 14, 15]. We will see that the situation is not so simple. In general, one must also demand that the system exhibits no bias for any particular microstate, a point emphasized by Tolman. Indeed, Tolman [9, see Sect. 25, particularly, pp. 63-64] uses this property of a statistical system as a postulate, when he discusses the validity of statistical mechanics, as does Sethna [10, Sect. 5.3]. This postulate should be valid even for non-equilibrium states that appear in a system as we vary macroscopic conditions; otherwise, as shown below, the temporal entropy of the system will violate the second law even if the system remains in equilibrium. This fact has not been appreciated to the best of our knowledge. Accordingly, ergodicity is broken even when the system is in equilibrium and there is no irreversibility. This fact also makes the temporal entropy quite useless.

When a microstate has been identified by a microstate measurement, we will call it the initial microstate. Let us now follow the evolution in time of the system prepared in the initial microstate. For this, we imagine making $N_{\mathrm{S}}$ replicas, each prepared identically in the initial microstate $j_0$, so that the probability is given by (22). Now, if the system, i.e., its microstate evolves deterministically, such as when there exists a Hamiltonian to govern this evolution, then at each instant of time, the initial microstate $j_0$ will have evolved into a unique microstate. The Liouville theorem follows immediately from such a deterministic dynamics, which is then used to prove the constancy of equilibrium entropy, both temporal and ensemble [8, 9, 24]. Indeed, those who insist on ergodicity, see for example [9, p. 69] and [10, Chap. 4], consider the evolution of a thermodynamic system to be governed by a deterministic dynamics. The lack of a dynamics, such as when the system is frozen in a basin (or a metabasin, which we discuss below) is a special case of a deterministic dynamics.

The duration $\delta$ is the time for a microstate to change into a different microstate. It is well known that in a deterministic dynamics of a system confined to a *finite volume*, the sequence of microstates

$$j_0, j_1, j_2, \cdots, j_{W-1} \qquad (23)$$

forms a *cycle* with the next microstate being the initial microstate: $j_W = j_0$ [33]. An important property of a deterministic evolution is that the mapping

$$j_k \Leftrightarrow j_{k+1} \qquad (24)$$

is *one-to-one*, so that it can be inverted ($j_{k+1} \to j_k$). This property gives rise to *time reversibility* in the evolution. The time for the recurrence of the initial microstate is known as the Poincaré recurrence time $t_\mathrm{R} \equiv W\delta$ [24]. During this period, the probability distribution will not change: it is $p = 1$ for the microstate $j_0'$ into which the system has evolved at some later instant, and $p = 0$ for any other microstate. Thus, the *ensemble entropy* will remain zero at every instant in a deterministic evolution [34]. Since the entropy does not decrease, it does not violate the second law, even though the system is not in equilibrium, which requires the entropy to be maximum [7]. The temporal entropy, on the other hand, is determined by all the microstates that have appeared so far. It is clear from the cycle (23) that it will continue to rise from its initial value of 0 to its maximum value $\ln W$ during the first recurrence, as the system moves from one microstate to another until all $W$ microstates have appeared during this period so that $p = 1/W$ for any of the microstates. It would, however, be incorrect to conclude from this that the system is in equilibrium. To see this, we need to follow what happens after the first recurrence. During any later recurrence period, the microstates again evolve following the cycle (23), so that the probabilities first become *different* from $p = 1/W$, only to become $p = 1/W$ at the end of each cycle. Consequently, the temporal entropy will decrease from $\ln W$ and rise again to this value in a cyclic manner [35]; it will, however, never become zero. Hence, the ensemble and temporal entropies are different so that in the conventional sense the ergodicity is lost. However, the temporal entropy also violates the second law, while the ensemble entropy dose not, even though the system is not in equilibrium.

To overcome the objection that the above problems are because the system has been prepared in a very unusual state (initial entropy equal to 0), let us consider preparing the replicas in several *distinct* microstates (but not all the microstates), which we denote by $i_0, i_1, i_2, \cdots, i_w$, where $1 < w < W$. It is assumed that we have already performed a microstate measurement to know which microstate a replica is in. The microstates in the above list need not be sequential, though they are arranged to follow the cyclic order (23). The replicas may be chosen so that the probabilities $p_k$ of a replica to be initially in the particular microstate $i_k, k = 1, 2, \cdots, w$ do not have to be equal to $1/w$ each; however, they still satisfy the sum rule in (13). The initial ensemble entropy $S_0$ of the system is given by (15), and satisfies the constraint

$$S_0 \leq \ln w < \ln W.$$

For the sake of simplicity, we will assume $p_k = 1/w$ for each of the allowed $w$ microstates so that $S_0 = \ln w$. This situation will be relevant for a glass kinetically frozen in a metabasin with $w$ distinct basins with $W$ representing the $N_\mathrm{b}$ basins. Let us now follow the evolution of this particular system. In a deterministic evolution (23), any initial microstate $i_k$ evolves into a unique microstate, which we call $i'_k$. It is not hard to see that $p'_k$ for the evolved microstate $i'_k$ is the same as the initial probability $p_k$ of the initial microstate $i_k$. Thus, the ensemble entropy at any later time, no matter how long, is equal to the initial entropy $S_0 = \ln w$. This is the same as above for $w = 1$. On the other hand, the evolution of the temporal entropy depends on how many initial microstates are considered. If one takes only one replica in a particular microstate, then the initial entropy is zero. This situation has already been discussed above. However, if one takes all of the $w$ distinct microstates as in the above ensemble average, then the initial entropy is $S_0 = \ln w$. Indeed, one can take any number $1 \leq \nu \leq w$ of initial microstates. The final entropy reaches $\ln W$ in all cases during the first recurrence. One easily arrives at this conclusion by recognizing that each initial microstate $i_k$ follows the same cycle of microstates in (23) so that each one has the same recurrence period, during which each microstate generates all of the $W$ microstates once. Therefore, at the end of $t_{\rm R}$, each microstate has appeared exactly $\nu$ times, and the above conclusion follows. After that, it will show a cyclic behavior similar to the one found above for $w = 1$ due to the same reason as above that the probability of different microstates begin to deviate from $1/W$. Thus, ergodicity remains broken even with an initial state, which is not an atypical state. Note again that the temporal entropy violates the second law but the ensemble entropy does not. Moreover, this broken ergodicity has nothing to do with any phase transition or a glass transition, as we have not even specified any particular system except that its phase space contains $W$ distinct microstates and that it is governed by a deterministic dynamics [34]. One should contrast the cyclic behavior of the temporal entropy with that suggested by Zermelo [36], who had argued that the temporal entropy will revert to the initial entropy in each cycle. One only has to consider the $w = 1$ case considered above to be convinced of this distinction.

We thus see that for $1 \leq w < W$, the temporal entropy makes no sense because of the cyclic nature over Poincaré's recurrence since it violates the second law. The ensemble entropy is well defined and devoid of any pathological behavior.

> CONCLUSION *The most important observation to make is that there is no irreversibility here, see (24), even though the ergodicity (at infinite time) is lost. Moreover, over a finite period of time, there is no fundamental reason for the two entropies to be the same, especially since the temporal entropy is not uniquely determined.*

This conclusion is important as the glass transition at which the loss of ergodicity occurs (see ELC3) [2, 14, 15], is also the point where irreversibility comes into play [1, 15, (b)]. However, the above discussion shows that the loss of ergodicity is not always related to irreversibility. The origin of irreversibility has to be found outside of a deterministic dynamics in the form of stochastic dynamics; see below and [34]. Moreover, it should be emphasized that ergodicity is a limiting (infinite time limit) concept and has no meaning over a finite duration. Despite this, it has been applied to glass transitions [2, 14, 15] with finite $\tau$.

The ensemble entropy is equal to its maximum possible value $S_0 = \ln W$ only for $w = W$, so that the system is by definition [7] in equilibrium. To see if it is also ergodic, we need to consider the temporal entroipy. If and only if we take all the $W$ micorstates initially can we have the temporal entropy equal to $\ln W$, so that the ergodicity holds at all times. The choice $w = W$ is precisely what is required for an unbiased preparation; see Tolman [9]. The interesting point to note is that *the temporal entropy will not show the above-mentioned cyclic behavior*; it remains constant at $\ln W$ from the beginning. However, this situation is not relevant for the glass transition. For the latter, the relevant situation corresponds to $1 \leq w < W$, when the system is kinetically frozen into a basin with $w$ basins. But for this situation, the temporal entropy and ergodicity make no sense as noted above. Thus, in this case, *not only the ergodicity is lost even though the system is in equilibrium but the temporal entropy also violates the second law.*

The above example clearly shows the importance of unbiased sampling or preparation we have discussed above.

The application of the above discussion to glasses requires another look because of the absence of any dynamics as the glass is kinetically trapped in a basin or metabasin. The same situation also occurs for the non-interacting Ising system discussed earlier. The above scenario will then change for the temporal entropy (it is this entropy that requires a dynamics to make sense), since there is no chance for the initial microstate to evolve into another microstate for $t < \tau$. Absence of any dynamics is no different from a deterministic dynamics. Thus, irreversibility is not an issue again as above. The only difference is that $\delta$ becomes too large compared to the experimental time scale $\tau$: $\delta >> \tau$. As the microstates do not change, no new microstates emerge from them in time. Therefore, the ensemble entropy remains constant at $S_0 = \ln W$. The temporal entropy, however, behaves very differently depending on the number $\nu$ of microstate that is considered initially. In general, we expect $\nu \leq w$, the number of basins in a metabasin. The entropy remains constant at the initial entropy $S_0 \equiv \ln \nu$. The ergodicity is lost for all $1 \leq \nu < W$ while it remains intact for $\nu = W$ for $t \leq \delta$, even though the system is in equilibrium in all cases. The loss of ergodicity ($\nu \neq W$) again does not imply any irreversibility; see the discussion above. This dependence of the temporal entropy on the choice of the initial collection of microstates is unsettling in that it is not clear how many microstates should be considered initially. The choice of $\nu = W$ is preferable because it gives rise to ergodicity if the latter is considered desirable. But this choice will not make any sense when the system is kinetically frozen in a metabasin ($w < W$). It is interesting to note that the system does not violate the second law for $t \leq \delta$, as both entropies remain constant. This is despite the fact that the microstate measurement has been performed, which allows us to prepare the system with identified microstates ($v \leq w < W$). A system prepared in an experiment, without the intervention of the microstate measurement will not have the *bias* to pick a particular metabasin with its own set of metabasins; see Tolman [9, see Sect. 25, particularly, pp.63-64]. For such a system, we have no information as to which of the metabsins or the basins the system is in. The lesson to be learned is that the microstate measurement destroys the statistical nature of a real system. But if such a measurement has not been made, the statistical nature of the system will keep it unbiased, so that we need to consider all possible groups of metabasins involving all $W$ basins. This is what Tolman calls an appropriately chosen ensemble [9, see p. 64].

The above discussion should leave no doubt in anyone's mind that the temporal entropy being non-monotonic cannot be considered a viable candiadate for the entropy of a system obeying deterministic dynamics, a kinetically trapped glass being one such system.

A *real* system will always have stochastic interactions with the surroundings that will result in transitions among microstates so that eventually we lose information about the initial microstates $i_k$ as they evolve in time and eventually the entropy reaches its maximum value of $\ln W$ that the system had just before the microstate measurement. What the above discussion, especially regarding the temporal entropy, illustrates is the importance of stochastic interactions, no matter how weak, with the surroundings [34] in inducing transitions among microstates so that the behavior of the system becomes consistent with the second law. There is no Poincaré recurrence now. Without any stochasticity, the evolution of the system becomes deterministic for which the ensemble entropy remains constant [34] as discussed above, but the temporal entropy ususally oscillates. Therefore, in the following, we will always consider the evolution of a system to be *stochastic*, unless noted otherwise.

### E. Causality

We now visit the issue of causality raised by Kivelson and Reiss [15]. Causality in its standard form implies a one-to-one relationship between cause and effect. In mechanics, it refers to the unique dynamical evolution of a mechanical system due to deterministic evolution, as shown in (23): its state $j_0$ at time time $t_0 = 0$ evolves uniquely to the state $j_1$ at time $t_1 = \delta > t_0$, which then evolves uniquely to the state $j_2$ at time $t_2 = 2\delta > t_1$, and so on; see [37, Chapter Two; see also Sect. 9.3.1 in [38]]. However, this one-to-one relationship between the cause ($j_0$) and the effects ($j_1, j_2, \cdots, j_K, j_{K+1}, \cdots$) is valid only if the space is unbounded, where we can state categorically that $j_0$ causes $j_K$, but $j_K$ cannot cause $j_0$: the present can only be affected by the past, but not by the future. In a finite volume, which is what a thermodynamic system will have, the existence of Poincaré's recurrence [24] destroys this causality, as $j_1$ will cause a microstate $j_R = j_{W-1}$ to emerge immediately prior to the recurrence; this microstate will then cause $j_0$ at the next instance. Now, causality as a principle needs to be validated by repeated experiments, just as thermodynamics needs to be. Thus, we need to verify causality in different trials. We begin to notice the problem when we do it. We will find that in some of the samples, $j_0$ will be the cause and $j_R$ will be the effect, while in other samples, $j_R$ will be the cause and $j_0$ will be the effect. The causality has been, strictly speaking, lost. We concede that observing the evolution $j_0 \to j_R$ will take unusually long time, but that is not the issue. The issue is whether causality can be treated as an *infallible* principle on which to base a theory. The above discussion shows that it is not so for a deterministic system when it is confined to a finite volume such as a thermodynamic system.

However, a thermodynamic system is not purely deterministic; rather, it is stochastic. For a stochastic dynamics, the issue of causality appears to have no role because of the probabilistic nature of its evolution [34]. A microstate $j_0$ at time $t_0$ evolves into many microstates $j_{01}, j_{02}, \cdots, j_{0k}$ at time $t_1 > t_0$ with probabilities $p_{01}, p_{02}, \cdots, p_{0k}$; each of these microstates $j_{0i}$ then evolves into microstates $j_{0i1}, j_{0i2}, \cdots, j_{0il}$ at time $t_2 > t_1$, with probabilities $p_{0i1}, p_{0i2}, \cdots, p_{0il}$; and so on. It is highly likely that one of the $j_{0im}$ is the microstate $j_0$, which makes the idea of causality quite irrelevant, as far as the causal relation among microstates is involved. As in the above deterministic evolution, some samples will show $j_0 \to j_{0im}$, while others will show $j_{0im} \to j_0$, so the concept of causality is meaningless.

One can still talk about causality with respect to the phase-space volume, as discussed by Penrose [29, p. 1941]. The causality now means that the phase-space volume, to be precise, the number of microstates $w(t)$ at some instance $t$ can only be affected by the past, and not the future. This concept of causality is much weaker than the original concept of causality related to the microstates, since the phase-space volume $w(t)$ actually keep track of their number only and not the individual microstates. Now, for a system in which the dynamics is absent or is frozen (as in a glass), then $w(t) = N_b$ for the number of basins is most certainly determined by the past in that it is determined by the rate of cooling, and is not determined by the future. Thus, the causality in the sense of Penrose is maintained in the glass.

The following daily life example will probably clarify the idea better. Consider throwing a die in a cup so that the die is always concealed. We shake the cup with the die vigorously and put it on a table face down so that we cannot see the die. Obviously, the outcome is one of the six faces of the die, but which one we do not know. The phase-space in this case consists of the six outcomes. Now, there is no one who believes that once the die has landed, it would on its own jump and change the outcome. In other words, the die once it has landed can never access the other outcomes, no matter how long we wait, unless we intentionally disturb it. At the same time, we all agree, assuming that the die is not loaded, that the probability of an outcome, a state (outcome) property, is $1/6$. The state property is determined by the collective set of outcomes, the phase-space volume $W = 6$, even though Reiss suggests that the principle of causality forbids the probability of the outcome to be affected by other outcomes which it is "unaware" of. This is obviously wrong as the value of the probability of any outcome is determined by the total number $W$ of *independent* outcomes.

One must not take the independence of different systems in the ensemble, which is a prerequisite in the ensemble approach, see our FUNDAMENTAL AXIOM, to imply that the probabilities of different outcomes (microstates) are also independent. They are two different concepts. In the first place, the definition of the probability, see (13), requires knowing how many (independent) microstates are there. This definition gives rise to the sum rule in (13), so that the probabilities of microstates are influenced by all the systems in the ensemble. This is not a violation of causality; it is merely a reflection of how the probabilities have to be defined: the sum rule creates a subtle correlation as far as the probabilities are concerned; despite this, the systems remain independent. There is no paradox here. Despite the independence of the systems, the entropy is given by (15), which is an average quantity and has to be averaged over the ensemble per our FUNDAMENTAL AXIOM. It is not a property of a single system.

Let us now apply the above idea to a kinetically frozen glass. We prepare $N_\mathrm{b}$ replicas, each replica representing a kinetically frozen system in one of the distinct basins, so that the probability for the system to be frozen in a basin $\alpha$ is exactly $\mathfrak{p}_\alpha = 1/N_\mathrm{b}$. This is despite the fact that each replica represents an independent sample so that no replica can be aware of other replicas, an argument that has been made by Kievelson and Reiss [15]. They use this argument to suggest that the entropy of the glass is zero. However, the ensemble entropy requires an average over all microstates [33]. Therefore, the initial ensemble entropy is $S_0 = \ln N_{\rm b}$. Let us now observe the evolution of the replicas in time. It is clear that each replica evolves independently, so that there is no causal relationship between any two of them. Indeed, none of the replicas change as there is no leaving of the basin. The independence of the replicas does not forbid the probability of a given replica to be determined by all other replicas. Causality in the sense used by Reiss (and not due to Penrose) and probability are two distinct concepts, and this distinction should be kept in mind. It is clear that at any time during the evolution, all replicas remain independent; despite this, the ensemble entropy remains non-zero ($= \ln N_{\rm b}$) and constant [34]. This constancy follows immediately from Liouville's theorem in classical or quantum mechanics, according to which the density or the density operator remains invariant.

What the above discussion shows that the just because the microstates at any given instance are independent does not mean, in contradiction to what Kievelson and Reiss [15] suggest, that the non-zero entropy violates their sense of causality. This causality has nothing to do with the value of the entropy.

## V. GLASS TRANSITION

### A. Residual Entropy

After a careful analyses of central issues at the heart of the controversy, we are ready to proceed with the issue raised by Goldstein [1]. Let a supercooled liquid, which can freeze into one of macroscopically relevant $N_{\rm b}$ *distinct* basins (after taking permutation and other symmetries into account) at the experimental glass transition temperature $T_{\rm g}$ and pressure $P$. We will call the resulting frozen structure of the supercooled liquid in a basin a *glass form* to distinguish it from the glass which represents the macrostate of the system $\Sigma$ for reasons that will become clear soon. These glass forms are indicated by an index $\alpha = 1, 2, \cdots, N_{\rm b}$. It suffices to say at present that the glass forms are the analogue of "microstates" for the purpose of calculating the residual entropy, even though the former are functions of the external variables but conventional microstates are not, and the glass is the analog of a "macrostate" of the system formed out of these "microstate." This analogy provides us with a hint to the resolution of the paradox. The nature and the number of basins or glass forms is controlled by the experimental time scale $\tau$ and can be manipulated by varying $\tau$. In the following, we will keep $\tau$ fixed as the temperature is lowered. Thus, $N_{\rm b}$ is a function of $T_{\rm g}(\tau), P$, and $N$. These glass forms (basins) have the same Gibbs free energy $G_{\rm b}(T, P, N, t \geq \tau)$ at $T_{\rm g}$, so that they are equally probable to occur (to be occupied) as the probability for the basin $\alpha$ is given by

$$\mathfrak{p}_\alpha = \mathfrak{p}_{\rm b} \equiv A \exp(-G_{\rm b}(T, P, N, \tau)/T), \quad T \geq T_{\rm g}, \tag{25}$$

and is the same for all basins; here $A$ is a normalization constant to ensure that the probabilities over all forms add up to unity. It follows from this that

$$\mathfrak{p}_{\rm b} = 1/N_{\rm b}.$$

This form of the probability is easily justified for $T \geq T_{\rm g}$. As the supercooled liquid is not a frozen state for $t = \tau$, all the basins that are explored by it are explored with no bias. Hence, their free energies must be the same at all times $t \geq \tau$. The form (25) follows from equilibrium statistical mechanics applied to the $NTP$-ensemble. We now argue that the equiprobability feature remains valid also at lower temperatures. For this, we make a large number of replicas of the system $\Sigma$ at $T_{\rm g}$ so that all glass forms have the same probability $\mathfrak{p}_{\rm b}$. At lower temperatures, each replica will remain frozen in the glass form it was in at $T_{\rm g}$. This freezing then ensures that the probability of each form remains unchanged from $\mathfrak{p}_{\rm b} = 1/N_{\rm b}$. However, it should not be concluded from this result that the Gibbs free energies of all basins are also equal, since (25) is no longer applicable below $T_{\rm g}$ where the system in not in equilibrium.

A more direct way to appreciate this aspect of the basin Gibbs free energy is to recognize that the $N_{\rm b}$ basins need not be identical in their shapes and depths, that is their topological properties need not be the same. There are no general arguments to suggest that their bottoms will have the same enthalpies. Indeed, it is highly likely that there is a distribution of energy and volume minima for the $N_{\rm b}$ basins. However, as we will see later, this observation is not relevant for our discussion here, which only deals with the residual entropy. Despite different depths from their common enthalpy at $T_{\rm g}$, all glass forms will be at the bottom of their corresponding basins at absolute zero. Thus, over the same temperature range from $T_g$ to absolute zero, each glass form will come down to reside at its respective bottom. Hence, the Helmholtz free energies of all glass forms are not going to be the same for $T < T_g$. Similarly, their Gibbs free energies also have no reason to be equal. We can still speak of an average Gibbs free energy $G_b(T, P, N, \tau)$, an average Helmholtz free energy $H_b(T, P, N, \tau)$, and an average basin entropy $S_b = -(\partial G_b/\partial T)$ for the glass over all glass forms for all $T < T_g$. All this follows from our FUNDAMENTAL AXIOM.

The equiprobability $\mathfrak{p}_b = 1/N_b$ aspect for various glass forms in no way implies that the system jumps from one basin to another during the period $\tau$; see REMARK in Sect. IV A. Exploration of different basins from a given basin will only occur for time period longer than $\tau$, which is not allowed in the present context. Hence, the freezing of the system into different glass forms makes the glass very different from the metastable supercooled liquid. There is no dynamics in the system anymore for $t < \tau$, so that we are dealing with a special case of a deterministic dynamics. As said earlier, the temporal entropy being non-unique makes no sense for finite times. Accordingly, we consider the ensemble entropy so as to conform to the principle of reproducibility..

For this, we apply the ensemble idea to different glass forms by treating each as a distinct "microstate." As these glass forms are time-dependent, they are not truly microstates in the conventional sense, since the latter are independent of time and external fields as said earlier. Thus, our approach is a generalization of the standard ensemble approach to time-dependent "microstates." The entropy arising from these glass forms represents the residual entropy. (In contrast, if we consider the microstates of the system, then their number will give us the entropy $S$.) To calculate this entropy, we use (15) to the glass forms by replacing $p_j$ by $\mathfrak{p}_\alpha$ and the sum over $j$ by $\alpha$ or use (17) and replace $W$ by $N_b$. We find that the residual entropy of $\Sigma$ is thus given by

$$S_R(T_g, P, N) = \ln N_b(T_g, P, N); \qquad (26)$$

Then, the free energy of system $\Sigma$ in the glassy state, the macrostate, will be given by

$$G(T, P, N, t) = G_b(T, P, N, t) - T S_R(T_g, P, N). \qquad (27)$$

From what has been said above, the freezing of the system in various glass forms does not change their probabilities. It follows therefore that neither the residual entropy nor the number $N_{\text{b}}$ of basins change for all $T \leq T_{\text{g}}$ for a fixed observation time $\tau$. Neither of the two will be true if $\tau$ does not remain fixed. The entropy of $\Sigma$ is given by

$$S = S_{\text{b}} + S_{\text{R}}, \qquad (28)$$

where $S_{\text{b}}$ is the intrabasin entropy introduced above and represents the average (over glass forms) entropy of a glass form. The component $S_{\text{b}}$ is usually identified as the vibrational entropy, but this is a misnomer since there are no vibrations inside any basin as the kinetic energy has been separated out of the configurational partition function. This entropy gradually vanishes as $T$ approaches absolute zero, and one is left with the residual entropy which represents the lack of information about which basin or glass form a given sample is frozen in.

If the glass transition results in the system kinetically freezing in a metabasin, then the configurational entropy at the glass transition will be obtained by replacing $N_{\text{b}}$ by $N_{\text{MB}}$ in (26). The metabasin entropy will then contain the average (over all metabasins) configurational contribution from the number of basins in the metabasins.

### B. Residual Entropy of Subsystems

#### 1. Reduction of probability Approach

It is possible to justify ELC by assuming that the kinetically frozen glass in one of the basins, which we label $\alpha_0$, changes the probability form $\mathfrak{p}_{\text{b}} = 1/N_{\text{b}}$ to $\mathfrak{p}_{\alpha_0} = 1$, and to $\mathfrak{p}_{\alpha'} = 0$ for all other basins $\alpha'$. Then the use of (15) immediately gives a zero configurational entropy, which then can be used to justify the entropy loss by $S_{\text{R}}$. Such an entropy reduction was found to occur due to a microstate measurement, see Sect. IV D. While quite an appealing argument for the justification, it overlooks two important facts:

1. In experimental glass transition, no such measurement is ever made that identifies precisely which basin the glass is frozen in. Such a measurement will tell us precisely the positions of all the $N$ particles which allows us to decipher which particular glass form the glass is in.

2. Because of the lack of such a measurement, we must determine the entropy by averaging over all glass forms; hence, the probability to be in any one of the basins remains exactly $\mathfrak{p}_{\text{b}} = 1/N_{\text{b}}$ ensure an unbiased situation.

We illustrate our point by a simple dice game using a single die in a cup. The dealer shakes the cup vigorously and puts it on the table so that we cannot see the die. We consider only six outcomes for the die, so that the phase (state) space for the die contains six microstates. We assume that the die is not loaded. Then, each outcome is equally probable. In that case, we can use (17) to calculate the entropy, which is $S = \ln 6$. This is also what one obtains from using (15), as the probability of each face is $p = 1/6$. This ensemble entropy remains constant forever, as long as the cup is not removed to reveal the die. The temporal entropy also remains constant and equal to the ensemble entropy ($\ln 6$). Now, not knowing what face will be on the top side of the hidden die, you bet some money that it is 3. At the time of the bet, the probability $p_3$ to get 3 is $1/6$, assuming the die to be not loaded. Your chances of winning is $1/6$, which is reflected in the value of the entropy. Indeed, at this moment, the probability $p_j$ to get any outcome $j = 1, 2, \cdots, 6$ is exactly $1/6$ due to the equiprobability assumption (not loaded). As soon as the cup is removed to reveal the die and the outcome $j_0$ (which is analogous to performing a microstate measurement), the probability suddenly changes to 1 for the outcome $j_0$, and to 0 for all other outcomes. The outcome $j_0$ means that no other events are possible. The consequence of the outcome is the following. If the outcome is $j_0 = 3$, you win the bet with *certainty*. If the outcome is different, you certainly lose the bet. The outcome also changes the phase space. It only contains one microstate corresponding to $j_0$ at the moment $t = 0$ when the cup was removed. The remaining five microstates are no longer there. In other words, as long as the die is concealed, the phase space contains six microstates and the probability of each microstate is $1/6$. This is similar to what happens for a glass, which can be in any of the basins so that If a which suddenly reduces discontinuously when the outcome is known. If and only if a microstate measurement has been performed that the phase space suddenly changes to have only one basin reflecting the effect of the measurement.

2. *Situation with a glass form*

The same situation is with the glass. All we know for sure is that the glass is in any one of the $N_{\text{b}}$ glass forms. We do not know the actual glass form it has. This situation is identical to the die which is known to be in one of the six possibilities when it is concealed by the cup, or to the non-interacting Ising spins discussed in Sect. IV A. The phase space contains all the basins to reflect this situation. The information that the glass has been formed is most certainly not equivalent to knowing precisely the particular glass form in which the glass is trapped. The latter will require the phase space to contain only one, the particular microstate. Whether there are transitions between different glass forms over the short duration $\tau$ is not relevant; see REMARK in Sect. IV A. The residual entropy is obtained by taking the average over all the glass forms or samples, and not only one glass form in which a sample may be. This is in accordance with the FUNDAMENTAL AXIOM. This leaves the residual entropy at its value $S_\text{R}$ given by (26). This is consistent with the already established rigorous CONCLUSION in Sect. III that the entropy cannot be discontinuous.

### 3. Residual Entropy of a Subsystem

The conclusion of the no loss of residual entropy is also consistent with the additivity principle. We imagine cutting the system $\Sigma$ into several equal parts of size $N'$, these parts representing a smaller system $\Sigma'$. Each part must represent a glass form or basin for the smaller system $\Sigma'$ in which the part must be kinetically frozen. The number of distinct glass forms or basins $N'_\text{b}$ for $\Sigma'$ is obviously

$$N'_\text{b} = (N_\text{b})^{N'/N}.$$

Again, as we cannot be sure of which glass form each part will represent, this ignorance then results in a non-vanishing residual entropy of

$$S'_\text{R} = \frac{N'}{N} \ln N_\text{b} \qquad (29)$$

for $\Sigma'$, as discussed above. As there are $N/N'$ parts in $\Sigma$, the residual entropy of $\Sigma$ is $\ln N_\text{b}$ as expected. The additivity principle has been *restored* by a careful analysis.

It is evident now that to conclude that the residual entropy has vanished just because a sample has frozen into a single basin or glass form is incorrect. The entropy reduction only happens if a microstate measurement is performed to identify the particular basin the glass is in. This discussion also shows that calorimetric measurements explore different glass forms associated with these subsystems, which is also the conclusion drawn by Goldstein; see (34) and (35) below.

## VI. INTERNAL EQUILIBRIUM AND RELAXATION BELOW THE GLASS TRANSITION

At the glass transition at $T_\mathrm{g}$, the free energy is continuous. For $T \geq T_\mathrm{g}$, SCL is allowed to probe all disordered microstates (ordered crystalline microstates are excluded from the probe) of the average enthalpy $H(T_\mathrm{g})$. Consider $\Sigma_0$ and its various parts representing the replicas of $\Sigma$. The disordered microstates of $\Sigma$ associated with this enthalpy partition into *disjoint* groups, each group belonging to a particular basin. As the system is in equilibrium, each part $\Sigma$ is able to explore all of these groups during the period $\tau$ as it jumps from one basin (group) to another. There is no GL yet for $T \geq T_\mathrm{g}$. The glass will emerge only when we lower the temperature by some $\Delta T > 0$ below $T_\mathrm{g}$. At the lower temperature, the relaxation time of SCL will be longer than $\tau$ so that each part $\Sigma$ can only probe one of the disjoint groups of microstates during the period $\tau$. This results in different parts of $\Sigma_0$ represent different glass forms. This only occurs for $\Delta T > 0$. At $T_\mathrm{g}$, there is no difference in SCL and GL, just as there is no difference between a liquid and its vapor at the critical point; the free energy, enthalpy and entropy are all *continuous* at $T_\mathrm{g}$.

For $\Delta T > 0$, different glass forms will relax if we wait longer than $\tau$. The presence of relaxation implies that none of the forms are in equilibrium with the surroundings, the latter being at temperature $T$ and pressure $P$, after it has been cooled at constant pressure $P$ from some initial temperature $T' = T + \Delta T$ of the surroundings to $T$ of the surroundings; however, each part is supposed to be in internal equilibrium and will in general have a different temperature and pressure from the surroundings. For simplicity, let us focus on one such sample of $\Sigma$ in contact with the surroundings. The arguments below can be applied to all parts of $\Sigma_0$ simultaneously with slight modification, which we will not do here. Let us assume $\Sigma$ and the surroundings form an *isolated* system whose total energy, volume and the number of particles are denoted by $E_0$, $V_0$, and $N_0$, respectively, while that of the surroundings by $\widetilde{E}$, $\widetilde{V}$, and $\widetilde{N}$. The surroundings is considered to be very large compared to $\Sigma$. Then

$$E_0 = E + \widetilde{E}, \quad V_0 = V + \widetilde{V}, \quad N_0 = N + \widetilde{N},$$

with $E$, $V$ and $N$ referring to the system $\Sigma$. For the isolated system $\Sigma_0$, $E_0$, $V_0$, and $N_0$ are fixed. We will assume that the number of particles $N$ of the system is also fixed, which means that $\widetilde{N}$ is also fixed. However, the energy and volume of the system fluctuate. The entropy $S_0$ of the isolated system can be written as the sum of the entropies $S$ of the system and $\widetilde{S}$ of the surroundings:

$$S_0(E_0, V_0, N_0, t) = S(E, V, N, t) + \widetilde{S}(\widetilde{E}, \widetilde{V}, \widetilde{N}, t).$$

The correction to this entropy due to the weak stochastic interactions between the system and the surroundings has been neglected, which is a common practice. As the system is very small compared to the surroundings, we can expand $S_0$ in terms of the fluctuating quantities of the system

$$\widetilde{S}(\widetilde{E}, \widetilde{V}, \widetilde{N}, t) \simeq \widetilde{S}(E_0, V_0, \widetilde{N}, t) - \left(\frac{\partial \widetilde{S}}{\partial E_0}\right) E(t) - \left(\frac{\partial \widetilde{S}}{\partial V_0}\right) V(t).$$

The fixed $T$ and $P$ of the surroundings determine the two derivatives in the above equation

$$\left(\frac{\partial \widetilde{S}}{\partial E_0}\right) \simeq \left(\frac{\partial \widetilde{S}}{\partial \widetilde{E}}\right) = \frac{1}{T}, \quad \left(\frac{\partial \widetilde{S}}{\partial V_0}\right) \simeq \left(\frac{\partial \widetilde{S}}{\partial \widetilde{V}}\right) = \frac{P}{T},$$

and $\widetilde{S}(E_0, V_0, \widetilde{N})$ is a constant, independent of the system. The approximation in the above equation is valid due to the smallness of the system relative to the surroundings. Thus,

$$S_0(t) - \widetilde{S}(t) = S(t) - H(t)/T = -G/T, \tag{30}$$

where we have suppressed exhibiting unnecessary quantities and where

$$G(t) = H(t) - TS(t)$$

is the time-dependent Gibbs free energy of the system $\Sigma$ with the surroundings at fixed $T$ and $P$. We remark that $T$ should not be confused with the time-dependent temperature $T(t)$ of the system. As this sample still possesses the residual entropy as discussed above, we can think of it to represent the system $\Sigma$. We further assume that the sample or the system is not in thermal equilibrium but is in mechanical equilibrium with the surroundings, merely to simplify the discussion below. Hence, its pressure is the same as that of the surroundings but its temperature $T(t) \leq T$ changes with time until the system reaches equilibrium when its temperature becomes equal to the surrounding temperature $T$.

We now consider this glass for a longer duration than $\tau$ to investigate its relaxation behavior. As there is internal equilibrium at each instant in the glass, its entropy can be treated as a function of $H(t) \equiv E(t) + PV(t)$ and $P$; there is no explicit $t$-dependence. It then follows from the identity $dH(t) = T(t)dS + V(t)dP$ that

$$(\partial S/\partial H)_P = 1/T(t). \tag{31}$$

The relaxation that occurs in the system $\Sigma$ originates from its tendency to come to thermal equilibrium in which its temperature $T(t)$ moves towards the external temperature $T$ from above; recall that we are considering a cooling experiment. This temperature change brings about a change in its enthalpy. Thus, the value of the entropy changes with time because the enthalpy changes with time.

The relaxation process results in the lowering of the corresponding Gibbs free energy, which results in not only lowering the enthalpy, as observed experimentally during aging, but also the entropy $S$ during relaxation. To demonstrate this, we proceed as follows. The time variation of the Gibbs free energy $G(T, P, N, t) = H(t) - TS(t)$ can be thought of as a variation due to the enthalpy. Therefore,

$$dG/dt = (\partial G/\partial H)_{T,P}(dH/dt) = (dH/dt)[1 - T/T(t)] \leq 0,$$

where $(\partial G/\partial H)_{T,P}$ denotes the derivative at fixed $T$ and $P$ of the surroundings (and not at fixed $T(t)$ and $P(t) = P$ of the system). It follows from this equation that

$$(dH/dt) \leq 0. \tag{32}$$

This behavior of $H(t)$ is consistent with experimental observations and justifies our assumption that $T(t) \leq T$. From (31), we immediately conclude that

$$(dS/dt) \leq 0, \tag{33}$$

as shown in Fig. 3. This finally proves our statement from which it follows that ELC8 is incorrect. The lowering of $S$ may lower both components of the entropy in (28). This lowering of the entropy of the system in this case is not a violation of the second law, whose statement in the form of the law of increase of entropy is valid only for an isolated system in which the energy is kept fixed. In other ensembles, it is the corresponding free energy that decreases with time. This is seen clearly from (30), which shows that it is the difference $S_0(t) - \widetilde{S}(t) = -G/T$ that increases with time, or the Gibbs free energy $G(t)$ decreases with time. The relaxation of the system in the present case does not mean that its entropy must increase.

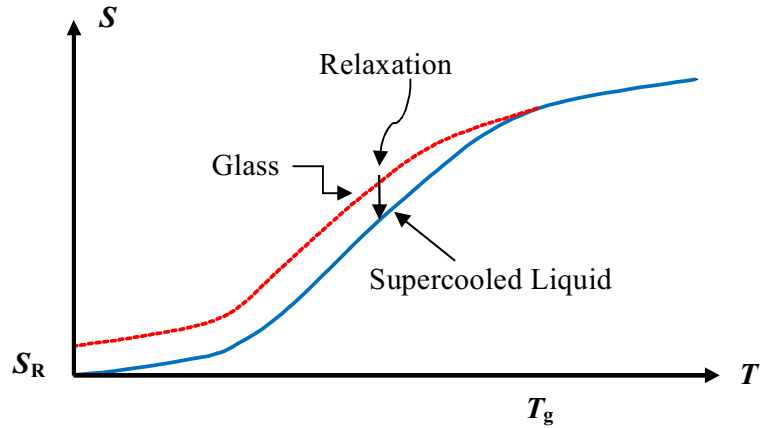

FIG. 3: Schematic behavior of the entropy for SCL (blue curve) and GL (red dotted curve). The GL entropy decreases, shown by the downward arrow, as it isothermally (constant temperature $T$ of the surroundings) relaxes towards SCL, during which its temperature $T(t)$ also decreases towards $T$ of the surrounding. At absolute zero, the GL entropy is equal to $S_R$, while that of SCL vanishes. The ideal glass transition in SCL is not shown here.

It is instructive to compare the specific heat of the glass with the specific heat of the corresponding fully relaxed state obtained as $t \to \infty$. Let us assume that at time $t = 0$, we change the temperature of $\Sigma$ form some initial temperature $T'$ to $T \equiv T' - \Delta T, \Delta T > 0$ instantaneously. The initial enthalpy $H(0)$ is the enthalpy of the glass at temperature $T'$, and $H(\infty)$ the value of the enthalpy after complete relaxation at temperature $T$. We consider the system at $t = \tau$ and determine its enthalpy. The specific heat of the glassy sample at this instant is given by

$$C_{P,\text{g}} = \lim_{\Delta T \to 0} \frac{H(0) - H(\tau)}{\Delta T}. \tag{34}$$

Then, the corresponding specific heat after complete relaxation ($H(\infty) \leq H(\tau)$) is given by

$$C_{P,\text{relax}} = \lim_{\Delta T \to 0} \frac{H(0) - H(\infty)}{\Delta T} \geq C_{P,\text{g}}. \tag{35}$$

## VII. NERNST'S POSTULATE AND ERGODICITY

The existence of the residual entropy in no way violates Nernst's postulate; see Landau and Lifshitz [8, Sect. 64]. Real glasses are non-equilibrium systems. Therefore, there is no thermodynamic requirement that they be in their lowest enthalpy state at $T = 0$. Indeed, as said earlier in Sect. V A, different glass forms will most probably have different enthalpies at absolute zero; the corresponding glass, as shown in (14), will have its enthalpy given by their *average over all forms*. Correspondingly, there is no reason for their entropy to vanish at absolute zero. If the glass transition occurred in a way that ensures that there is only one distinct single basin in which the glass is trapped, then the glass is in a pure state as far as its configurational property is concerned. Accordingly, the residual entropy will certainly vanish, which will ensure that the entropy $S$ will vanish at $T = 0$. This will be consistent with Nernst's postulate. However, the necessary, though not sufficient, condition for this to occur is for the glass to be an equilibrium state. This will most probably happen in an *ideal experiment*, in which the supercooled liquid is cooled infinitely slowly (but still ensuring that it does not crystallize). At the *ideal* glass transition at $T = T_{\mathrm{K}}(P)$, the supercooled liquid is most probably trapped in this single basin so that the residual entropy $S_{\mathrm{R}}^{\mathrm{ideal}} = 0$ for an ideal glass. In real experiments constrained by experimental time limit $\tau$, one will obtain many distinct glass forms at $T = T_{\mathrm{g}}(P, \tau) > T_{\mathrm{K}}$, which will persist all the way to absolute zero and will result in a non-zero residual entropy.

This brings us to another confusion about real trapped glasses. Let us consider first $T = T_{\mathrm{g}}$, and prepare many samples or replicas of a GL (which is not different from SCL at $T_{\mathrm{g}}$ so no distinction needs to be made here), such that the probability of a GL to be in any of the basins is $\mathfrak{p}_{\mathrm{b}} = 1/N_{\mathrm{b}}$. All of these samples are characterized by the same enthalpy $H_0 = H(T_{\mathrm{g}}, \tau)$ at $T_0 = T(T_{\mathrm{g}}, \tau) \equiv T_{\mathrm{g}}$ and $P$. For time $t < \tau$, each sample remains kinetically frozen in its particular glass form. Despite this, the configurational (ensemble) entropy according to the FUNDAMENTAL AXIOM is determined by all the forms, which then yields $S_{\mathrm{R}} = \ln N_{\mathrm{b}}$. For $t \geq \tau$, the sample will jump from this glass form to others so that all the glass forms are explored but the probability of each glass form remains unchanged: $\mathfrak{p}_\alpha = \mathfrak{p}_{\mathrm{b}} = 1/N_{\mathrm{b}}$. Hence, the residual entropy remains equal to $S_{\mathrm{R}}$. There is no surprise here.

We wish to compare this entropy with the temporal entropy. For this, we need to first decide how many microstates or samples we start with initially. Let us first consider a single initial sample in one of the possible forms, the forms representing the "microstates." We have already established above that every sample of glass of $\Sigma$ possesses the residual entropy $S_{\mathrm{R}} = \ln N_{\mathrm{b}}$ so that its free energy is $G = G_{\mathrm{b}} - TS_{\mathrm{R}}$, see (29) and the discussion following it. However, for the calculation of the temporal entropy, we need to consider a sample in a "microstate" or glass form. Therefore, we need to identify the "microstate" of the sample by a microstate measurement, which we denote by $\alpha_0$. This measurement, see Sect. IV D, allows us to identify the locations of all the $N$ particles uniquely, which abruptly reduces the probability distribution from equal probability $\mathfrak{p}_\mathrm{b} = 1/N_\mathrm{b}$ for each form to

$$\mathfrak{p}_\alpha = \delta_{\alpha,\alpha_0},$$

where $\delta_{\alpha\beta}$ is the Kronecker delta. This gives 0 for the residual entropy right after the measurement, so that its free energy reduces to $G = G_\mathrm{b}$. We emphasize the following point: The microstate measurement should not be confused with usual measurements such as calorimetric measurements that do not identify the glass form. Therefore, the latter measurements do not raise the Gibbs free energy or lower the residual entropy. As the free energy has increased, the sample is out of internal equilibrium after this measurement. This glass form is still characterized by the same enthalpy $H_0 = H_{\alpha_0}(\tau) \equiv H(\tau)$ at $T_0 = T_{\alpha_0}(\tau) \equiv T_\mathrm{g}$ and $P$ even after the measurement. Despite this, the sample after the measurement is not in internal equilibrium as it is restricted to be in only one "microstate" or glass form. The internal equilibrium requires equiprobability distribution

$$\mathfrak{p}_\alpha = \mathfrak{p}_\mathrm{b} = 1/N_\mathrm{b}. \tag{36}$$

Now that we have prepared the sample in a particular form $\alpha_0$, we must follow the evolution of this "microstate" in time to evaluate the temporal entropy. However, we must ensure that the evolution occurs *adiabatically* at fixed $H_0$, $T_0$ and $P$ to any of the $N_\mathrm{b}$ "microstates." This should be compared with the temporal and adiabatic evolution of a microstate, for example in a microcanonical ensemble, which must occur at fixed $E, V$ and $N$ to microstates appropriately restricted so as to implement symmetry breaking, if any. After all, we wish to compare the temporal entropy with the ensemble entropy for a system at the same $H_0$, $T_0$ and $P$ confined to $N_\mathrm{b}$ basins. These conditions need to be implemented for a correct comparison. In the process no heat exchange should be allowed. The evolution then allows us to use (19) to calculate the entropy. As we watch this sample in time for $t \gg \tau$ (fixed $H_0$ at $T_0$ and $P$), the sample will make jumps to other glass forms at the same $H_0, T_0$ and $P$, until all glass forms are generated in time with equal probability $\mathfrak{p}_\mathrm{b} = 1/N_\mathrm{b}$, so that the temporal residual entropy eventually equals $S_\mathrm{R}$ and the Gibbs free energy becomes equal to $G_{\rm b} - TS_{\rm R}$. (There is no Poincaré recurrence as we are dealing with stochastic evolution.) This conclusion is the same as in Sect. V B that even a single sample has the residual entropy $S_{\rm R}$. The ergodicity in the infinite time limit sense is intact at $T_{\rm g}$.

It should be stressed that the temporal evolution that is considered above is stochastic in nature and not deterministic. Hence, the issue of recurrence is no longer relevant [34]. Thus, the temporal entropy will not suffer from the cyclic behavior observed earlier in Sect. IV D. In the following, we will assume the dynamics of evolution to be stochastic, and not deterministic.

Does the situation change below $T_{\rm g}$? It is clear that the ensemble average will still give the same residual entropy $S_{\rm R}$, since (36) is still valid as discussed in Sect. V A. Even the temporal residual entropy will eventually reach this value, as discussed above for $T = T_{\rm g}$. There are some issues that need to be addressed. The first issue we need to address is that we need to consider the time scale $\tau(T)$ over which a glass form will not jump to other glass forms, all belonging to the set of $N_{\rm b}$ basins in which the glass gets trapped. The other issue, which creates some complications, is that the enthalpies of all basins need not be identical. However, they have an average enthalpy $H(\tau)$, which needs to be maintained as the specified basin $\alpha_0$ begins to explore other basins for $t \gg \tau(T)$. The evolution of this basin must be constrained to maintain $H_0 = H(\tau)$, $T_0 = T(\tau)$ and $P$. This evolution will eventually explore all of the same $N_{\rm b}$ glass forms or basins with equal probability that were present at $T_{\rm g}$. This then ensures that the temporal and ensemble versions of the residual entropy become identical and equal to $S_{\rm R}$, and ergodicity remains intact. As long as $\tau(T) < \infty$, the above conclusion does not change. However, at very low temperatures below the Debye temperature, we will also have to worry about quantum tunneling between various forms [8], even if $\tau(T) < \infty$. This also does not affect our conclusion, which is clearly inescapable: Every sample of glass of $\Sigma$ possesses the same residual entropy $\ln N_{\rm b}$. Moreover, there is no difference between the two definitions of the entropy so that the ergodicity remains intact. The situation will not change if we take more than one initial "microstates" $\alpha_0$. The case when we take all $N_{\rm b}$ microstates is interesting in that the ensemble and temporal entropies are the same even for short time duration $t \geq \tau$.

Let us point an interesting aspect of the adiabatic relaxation of the above specially prepared glass form used for the temporal average, as it evolves in time. Since the evolution occurs at fixed enthalpy $H_0$, there is no heat exchange as the intrabasin equilibration occurs.

Because of this, such a relaxation does not affect the heat capacity of the glass. This *adiabatic relaxation* should not be confused with the relaxation considered in Sect. VI.

## VIII.  CONCLUSIONS

We have conducted a detailed analysis of some important concepts such as entropy, probability, ergodicity, irreversibility, slowly varying metastable states, causality, etc. whose relevance and interpretations have been questioned recently in the context of glasses. A central assumption in ELC is that of internal equilibrium in a glass [2, 14]. We accept this assumption in order to carry out our analysis, from which follow some important results. Because of internal equilibrium, we are forced to conclude that SCL does not undergo a discontinuous drop of $S_\text{R}$ in its residual entropy at the glass transition as required by ELC. The proof of this statement is quite general and does not depend on the idea of any landscape or any model calculation. It is based on the experimental observation that the enthalpy and volume are continuous and single-valued functions of the temperature and our insistence that the four fundamental principles of thermodynamics, see Sect. II, must also apply to slowly varying metastable states like glasses. Any sharp discontinuity, more than $\simeq 5\%$ of $S_\text{R}$ produces several inconsistencies with thermodynamics and experimental facts. We find that there is no way for SCL to turn into GL without either (i) violating the second law if GL is a non-equilibrium macrostate or (ii) requiring work or heat input from outside, which is inconsistent with experimental evidence, if GL is an equilibrium state; see Conclusion 2 in Sect. III. If the macroscopic discontinuity in ELC manifests itself in a continuous variation for a finite system in the transition region, then it must give rise to stability violation with the Gibbs free energy becoming convex with respect to the temperature in that range. The same violation will also occur if the entropy has a concave region near the transition region, as is the case with the numerical results of Mauro et al [14]. It follows from all these contradictions that the entropy loss conjecture must be rejected. If accepted, it leads to the violation of several thermodynamic principles and experimental facts, and will require a new thermodynamics.

We find that the basic premise of ELC is based on an incorrect interpretation of what it means to be confined to a basin. Confinement (unless we have a prior knowledge that there is only one basin for the confinement) itself does not mean that we have the complete information about the system. Accordingly, the residual entropy is not determined by the particular glass form. Rather, it is obtained by taking an average over all glass forms, as is the standard practice; see the FUNDAMENTAL AXIOM. Failure to adhere to this principle results in ELC. We find no justification for ELC9. Thermodynamics is the study of not a single system, which runs contrary to the what Gupta and Mauro claim; see their first quote in Sect. I A. It is the average description of many representative systems, an observation made by several workers in the field; see for example Gibbs [26], and Tolman [9]. This answers our first question Q1. The many systems (glass forms) alluded to above must in general represent *all* microstates. However, these microstates need not have to be accessed by the system during the observation time. It also follows from this observation that the ensemble entropy is more fundamental than the temporal entropy for glasses. It is the latter that gives rise to the impression that it is the number of microstates sampled by a system during its observation that determines the entropy. We give arguments to show that the temporal entropy does not make much sense, especially for glasses. Because of the finite observation time, this entropy is not suitable as it is not unique; see REMARK in Sect. IV B. Once this misunderstanding is corrected, we discover that the residual entropy is real and not fictional and this (ensemble) entropy will persist at absolute zero. Whether there is any dynamics is not relevant for the use of the ensemble entropy; the dynamics is required only if one is interested in evaluating the temporal entropy. The latter, as said above, is not relevant for glasses. This answers our second question Q2. Accordingly, the entropy is not determined by the number of microstates accessed by a system during its observation over a finite duration. This conclusion should be contrasted with the first quote of Gupta and Mauro in Sect. I A. This also means that the time average over the observation time will usually be different from the ensemble average; it is the latter, which according to the principle of reproducibility gives experimentally observed average or the thermodynamic average. A particular sample will usually show deviation from this observed or thermodynamic average; see the FUNDAMENTAL AXIOM. Thus, the behavior of a single sample should not be identified with the thermodynamic behavior, especially when the system is not in equilibrium such as a glass; see the second quote of Tolman in Sect. II.

There is no difference between GL and SCL at $T_g$ in direct contradiction with ELC2. This results in a continuous entropy, Gibbs free energy, etc. at the transition. Below the transition, GL and SCL represent two different macrostates. While SCL is in equilibrium with the surroundings, GL is not, although it is assumed to be in internal equilibrium. Accordingly, its thermodynamic properties change continuously until it equilibrates with the surroundings, when it becomes identical with SCL. The relaxation of a glass towards SCL results in the entropy decreasing with time along with its enthalpy. We find no justification for the entropy to increase in such a relaxation, which is in contradiction to ELC8. This is not a violation of the second law, since this relaxation is not adiabatic. While we do not believe ergodicity to be relevant for glasses as the concept is meaningful only we observe the system for infinitely long time [8], we still consider it as it has been generally invoked to describe glasses as a broken ergodic state. Almost all systems break ergodicity when we observe them for a short period of time, whether they form glass or not. Therefore, to worry about ergodicity in experimental glass transition makes no sense in our opinion. The only sensible way to check ergodicity is to consider infinite long time limit, which is what is required for its definition. We then find that it remains valid even in the glassy state, provided proper *adiabatic* condition is imposed on the evolution. This adiabatic condition is similar to what we impose on the evolution of an isolated system. This contradicts ECL3. The concept of causality proposed by Reiss [15] is based on the demand that different samples or glass forms be independent, which we argue that they are. However, we have argued that their independence is not inconsistent with the interdependence of sample probabilities or a non-zero residual entropy. Thus, there is no problem with causality as Reiss defines it. The causality introduced by Penrose [29] is also satisfied, even if the residual entropy is non-zero.

We thank Martin Goldstein for introducing us to the controversy and to him and Herman Cummins for fruitful communications. We thank Robin Speedy for informing us of [12], and Evguenii Kozliak for comments on an earlier version of the manuscript.

———